\documentclass[10pt,journal,doublecolumn]{IEEEtran}
\usepackage{threeparttable,placeins,makecell,stfloats,cite}
\usepackage{threeparttable,arydshln}
\usepackage{amsmath,amssymb,bm}
\usepackage[dvips]{graphicx}
\usepackage[usenames]{color}
\usepackage{amsfonts}
\usepackage{latexsym}
\usepackage{multirow}
\usepackage{caption}
\usepackage{graphicx}
\usepackage{subfigure}
\usepackage{rotating}
\usepackage{float}
\usepackage{url}
\usepackage{cite}
\usepackage[noend]{algpseudocode}
\usepackage{algorithmicx,algorithm}
\usepackage[noend]{algpseudocode}
\usepackage{diagbox}

\usepackage{amsthm}
\theoremstyle{plain}
\newtheorem{theorem}{Theorem}
\newtheorem{lemma}{Lemma}
\newtheorem{proposition}{Proposition}
\newtheorem{remark}{Remark}
\newtheorem{corollary}{Corollary}

\usepackage{graphicx}                 
\usepackage{subfig}                
\usepackage{overpic}  
\usepackage{enumitem}

\begin{document}

\title{Flag Sequence Set Design for Low-Complexity Delay-Doppler Estimation}

\author{Lingsheng~Meng, \emph{Graduate 
 Student Member, IEEE}, Yong Liang~Guan, \emph{Senior Member, IEEE},\\Yao~Ge, \emph{Member, IEEE}, and Zilong~Liu, \emph{Senior Member, IEEE}

\thanks{Examples of our proposed Flag sequences can be found in the link of \protect\url{https://github.com/meng0071/Flag_sequence}. This paper was presented in part at the 10th International Workshop on Signal Design and its Applications in Communications (IWSDA 2022) \cite{Meng2022}. \textit{ (Corresponding author: Zilong Liu.)}}
\thanks{Lingsheng Meng and Yong Liang Guan are with the School of Electrical and Electronic Engineering, Nanyang Technological University, Singapore 639798, and also the Continental-NTU Corporate Lab, Nanyang Technological University, Singapore 639798 (e-mail: meng0071@e.ntu.edu.sg; eylguan@ntu.edu.sg).}
\thanks{Yao Ge is with Continental-NTU Corporate Lab, Nanyang Technological University, Singapore 639798 (e-mail: yao.ge@ntu.edu.sg).}
\thanks{Zilong Liu is with the School of Computer Science and Electronics Engineering, University of Essex, Colchester CO4 3SQ, UK (e-mail: zilong.liu@essex.ac.uk).}
}

\maketitle

\begin{abstract}
This paper studies \textit{Flag sequences} for low-complexity delay-Doppler estimation by exploiting their distinctive \textit{peak-curtain} ambiguity functions (AFs). Unlike the existing Flag sequence designs that are limited to prime lengths and periodic auto-AFs, we aim to design Flag sequence sets of arbitrary lengths with low (nontrivial) periodic/aperiodic auto- and cross-AFs. Since every Flag sequence consists of a \textit{Curtain sequence} and a \textit{Peak sequence}, we first investigate the algebraic design of Curtain sequence sets of arbitrary lengths. Our proposed design gives rise to novel Curtain sequence sets with ideal \emph{curtain} auto-AFs and zero/near-zero cross-AFs within the delay-Doppler zone of operation. Leveraging these Curtain sequence sets, two optimization problems are formulated to minimize the weighted integrated masked sidelobe level (WImSL) of the Flag sequence set. Accelerated parallel partially majorization-minimization algorithms are proposed to jointly optimize the transmit Flag sequences and symmetric/asymmetric reference sequences stored in the receiver. Simulations demonstrate that our proposed Flag sequences lead to improved WImSL and peak-to-max-masked-sidelobe ratio compared with the existing Flag sequences. Additionally, our Flag sequences under the Flag method exhibit Mean Squared Errors that approach the Cramér-Rao lower bound and the sampling bound at high signal-to-noise power ratios.
\end{abstract}

\section{Introduction}
\IEEEPARstart{D}{elay-Doppler} estimation plays an important role in radar, sonar, navigation, and communication systems \cite{Ma2020, Liang2014, Kodheli2021,10891132}. In complex wireless channels with high mobility and multipath propagation, delay-Doppler estimation is often first carried out at the receiver in order to determine the relative distance and relative velocity of targets or to deal with the selectivity in both time- and frequency- domains \cite{Tse2005, Richards2005, Kumari2018}.

In radar sequence design theory, the ambiguity function (AF) is an important design metric that characterizes the receiver response for targets with different delays and Doppler shifts \cite{Skolnik2008, Rihaczek1967}. For sequence design of a single user, it is ideal to achieve an auto-AF (AAF) that consists of a spike at the origin and zero sidelobes elsewhere on the entire delay-Doppler plane \cite{Cook2012}. Achieving such an ideal AAF is impossible due to the limited volume of AAF \cite{Benedetto2009}. However, in practical scenarios, the maximum delay and maximum Doppler shift may be much smaller than the sequence duration and signal bandwidth, respectively \cite{Strohmer2003, Jing2019}. Thus, the optimization for spike-like AAF in certain delay-Doppler zone of operation (ZoO), also known as local ambiguity shaping, is of strong practical significance. Some efficient iterative algorithms have been developed for suppressing the weighted integrated sidelobe level (WISL) or the peak sidelobe level (PSL) of the AAFs with symmetric transmit sequence and receive reference sequence in \cite{Song2016,Arlery2016s, Cui2017}. On the other hand, the transmit sequences and the receive reference sequences can also be jointly designed in an asymmetric manner. By doing so, one can achieve increased design degree-of-freedom and lower sidelobes at the cost of a loss-in-processing gain (LPG) of the main AF peak \cite{Zhou2022, Rabaste2015}. In \cite{Wang2022b}, the LPG is effectively controlled by incorporating a peak constraint into the objective function as a penalty term. 

To improve the detection sensitivity and estimation accuracy, it is also critical to consider mutual interference between different radar stations or users in systems such as multi-static primary surveillance radar (MSPSR) and connected automotive radar systems \cite{Brooker2007}. This requires us to minimize the cross-AFs (CAFs) between multiple sequences. The iterative algorithms introduced in \cite{Arlery2016s} and \cite{Cui2017}, are respectively applied in \cite{Arlery2016m} and \cite{Liu2019} for efficient sequence set design with good CAF properties. Multiple coding schemes are proposed to minimize the CAF between two vehicles in \cite{Tang2018}. By extending the work in \cite{Tang2018}, a novel algorithm is developed in \cite{Bose2021} to mitigate mutual interference between multiple connected vehicles. In addition to the challenges in radar sequence design, integrated sensing and communication (ISAC) emerges recently as a new 6G use scenario by developing dual-function waveforms that meet both radar sensing and communication requirements \cite{liu2018muMIMO, sturm2011waveform}. For instance, prior works such as \cite{bazzi2023a, Huang2022, bazzi2023b} have proposed waveform designs for ISAC systems aiming for achieving low peak-to-average power ratios (PAPR) and reduced sidelobe levels.

In practice, low-complexity delay-Doppler estimation is highly desirable for lower processing latency, hardware storage size, and power consumption. For the aforementioned studies with spike-like AAFs, an exhaustive search in the delay-Doppler ZoO is often carried out to determine the delay and Doppler values of targets \cite{Skolnik2008}. This results in cubic computational complexity with respect to the sequence length or the size of the ZoO. Even with the aid of the fast Fourier transform (FFT), almost quadratic complexity is required \cite{Toole2010}. A remarkable breakthrough with almost linear computational complexity is reported in \cite{Fish2013} by leveraging the Flag property of Heisenberg-Weil sequences (HWS). Specifically, every HWS possesses a unique periodic AF consisting of a \emph{peak} and a \emph{curtain} (see Fig.~\ref{AF_ideal_HWS}(a) for an illustration). Such a property arises from the fact that every HWS is constructed by summing a pair of almost orthogonal Heisenberg sequences and Weil sequences. The periodic AAFs of Heisenberg sequences have \emph{curtain}-like shapes, while the periodic AAFs of Weil sequences are \emph{peak}-like. Sequences with such \emph{peak-curtain} periodic/aperiodic AAFs in the ZoO are called Flag sequences in this paper. Thanks to such \emph{peak-curtain} AAF property, an almost-linear-complexity delay-Doppler estimation process can be achieved by identifying the \emph{curtain} first, followed by the search of the \emph{peak}. 

Nevertheless, one major drawback of HWS is that they may suffer from high AAF sidelobes imposed by Weil sequences, as illustrated in Fig.~\ref{AF_ideal_HWS}(b). For those sidelobes located near the origin, detection ambiguity arises, and the accuracy of delay-Doppler estimation deteriorates. Another drawback of HWS is that some Heisenberg sequences may not have ideal \emph{curtain} AAF, resulting in performance loss. Furthermore, \cite{Fish2013} only addresses the construction of periodic AF Flag sequences, and the sequence lengths are limited to prime numbers only. 

In this paper, we propose a novel class of Flag sequence sets to address the aforementioned problems. 
The major contributions of this paper are summarized as follows:
\begin{enumerate}[]
\item We propose novel Flag sequence sets with \emph{peak-curtain} AAFs and low CAFs in the ZoO. Compared to the existing HWS that are only limited to prime lengths and periodic AF, our proposed approach can generate Flag sequences with lower AF sidelobes of arbitrary lengths for both periodic AF and aperiodic AF cases.
\item We first propose novel systematic construction of Curtain sequence sets by selective use of discrete chirp sequences. We prove that our proposed sequence sets possess ideal \emph{curtain} AAFs and zero/near-zero CAFs within the ZoO. These carefully designed Curtain sequence sets then enable the design of Flag sequence sets of arbitrary length.
\item We formulate two optimization problems aimed at minimizing the weighted integrated masked sidelobe level (WImSL) of Flag sequence sets with symmetric and asymmetric receive reference sequences, respectively. To solve these non-convex optimization problems, accelerated parallel majorization-minimization (AP-MM) algorithms are proposed. Our core idea is to transform the original problems into surrogate smooth problems in order to jointly optimize the transmit sequences and the receive reference sequences.
\item Extensive numerical results show that our proposed Flag sequence sets exhibit superior \emph{peak-curtain} AAFs in the ZoO compared to HWS, along with low CAFs. Additionally, our proposed sequence sets under the Flag method demonstrate improved detection and delay-Doppler estimation performances over HWS. Moreover, in high SNR scenarios, the mean squared errors (MSE) of our proposed Flag sequences under the Flag method approach the Cramér-Rao lower bound (CRLB) and the sampling bound (SB).
\end{enumerate}

The remainder of this paper is organized as follows. Section~II introduces some preliminary concepts and definitions. In Section~III, we present our Curtain sequence set constructions. In Section~IV, the Flag sequence sets design problems are formulated. Section~V presents the AP-MM algorithms for solving the formulated problems. In Section~VI, we present the simulation results to validate the effectiveness of the proposed Flag sequence sets. Finally, Section~VII concludes the paper.


\emph{Notations}: $(\cdot)^{\text{T}}$, $(\cdot)^{\dagger}$, $\text{Tr}(\cdot)$ and ${\text{vec} (\cdot)}$ stand for the transpose, conjugate transpose, trace and stacking vectorization of a matrix, respectively. $\text{Diag}(\bm \rho)$ is a matrix constructed with elements of $\bm \rho$ as its principal diagonal. $\left \| \cdot \right \| $ represents the 2-norm of a vector. $[\cdot]_N$ and $\left \lfloor \cdot \right \rfloor$ denote the modulo $N$ and round down operations. $(\cdot)^\ast $, $|\cdot|$, $\Re \left \{ \cdot \right \} $, $\Im \left \{ \cdot \right \} $ and $\text{arg}(\cdot)$ denote the conjugate, absolute value, real part, imaginary part and phase of a complex number, respectively. The imaginary unit is denoted by $j = \sqrt{-1}$. $\bm{I}_M$ and $\bm 0_{M \times  N}$ represent the $M \times M$ identity matrix and $M \times N$ zero matrix, respectively.
\section{Preliminaries}
In this section, we first introduce the delay-Doppler estimation schemes with symmetric/asymmetric reference sequences in the receiver with periodic/aperiodic AFs and discuss their computational complexity under the traditional exhaustive search method. Then, we introduce the Flag sequence and the associated low-complexity Flag method. 
\subsection{Periodic/Aperiodic AFs with Symmetric/Asymmetric Receive Reference Sequence}
\label{ZoODF}
The delay-Doppler estimation based on the AF is widely used for signal processing in various fields such as radar, sonar, and communication systems \cite{Verdu1998, Wang2022,meng2024bound}. Let us consider a transmit sequence $\bm s$ and a receive reference sequence $\bm r$:
\begin{subequations}
    \begin{align}
        \bm{s}={\left [s[0], s[1],\cdots,s[N-1] \right ]}^{\text{T}} \in \mathbb{C}^{N \times 1}, \\
        \bm{r}={\left [r[0], r[1],\cdots,r[N-1] \right]}^{\text{T}} \in \mathbb{C}^{N \times 1},
    \end{align}
\end{subequations}
where $N$ denotes the sequence length. The transmit sequence $\bm s$ and the corresponding receive reference sequence $\bm r$, are symmetric or matched when they are identical. Conversely, they are referred to as asymmetric or mismatched when they differ. It is important to note that \emph{symmetric} and \emph{asymmetric} refer to the relationship between a transmit sequence and its corresponding receive reference sequence\footnote{In radar theory, symmetric and asymmetric receive reference sequences are also referred to as matched and mismatched receive filters.}. Let $\tau$ be the delay shift bin and $\omega$ be the normalized Doppler shift bin. Then we define the discrete AF of a transmit sequence $\bm s$ and a receive reference sequence $\bm r$ with respect to $\tau$ and $\omega$ as \cite{Liu2019}
\begin{equation}
	\label{AFC}
	A_{\bm s,\bm r}(\tau,\omega)={\left |{\bm r}^{\dagger}{\bm J_{\tau}}{\bm s_{\omega}} \right |},
\end{equation} 
where $\bm s_{\omega}=\text{Diag}(\bm h(\omega))\cdot \bm s$ represents the frequency shifted version of sequence $\bm s$, with $\bm h(\omega)$ being the normalized Doppler shift vector defined by 
\begin{equation}
\bm h(\omega) = {\left [e^{j2\pi \omega/N},e^{j4\pi \omega/N}, \cdots,e^{j2 N\pi \omega/N} \right ]}^{\text{T}}.
\end{equation}
$\bm J_{\tau}$ is the $N \times N$ time shift matrix, for aperiodic AF, it is defined by 
\begin{equation}
	\label{shiftmatrixa}
	\setlength{\arraycolsep}{1.2pt}
	J_{\tau}[m,n]=\begin{cases}
	1, & n - m = \tau; \\
	0, & n - m \neq \tau.
	\end{cases} 
\end{equation}
For periodic AF, it can be represented as
\begin{equation}
	\label{shiftmatrixp}
	\setlength{\arraycolsep}{1.2pt}
	J_{\tau}[m,n]=\begin{cases}
	1, & [m - n]_N = \tau; \\
	0, & [m - n]_N \neq \tau.
	\end{cases} 
\end{equation}
Throughout this paper, we define the AF between a transmit sequence and its corresponding receive reference sequence of the same user as AAF, while the AF between a transmit sequence and a receive reference sequence of a different user is defined as CAF.

Furthermore, in practical scenarios, the delay-Doppler ZoO can be identified by considering the maximum target distance $d_{\text{max}}$, and the maximum relative velocity $v_{\text{max}}$. This ZoO can be expressed as a set:
\begin{align*}
    \bm \Gamma({\tau_{\text{max}},\omega_{\text{max}}}) \!= \! \left \{\! \left ( \tau,\omega \right )\!|\tau \!\in\![ -\tau_{\text{max}},\tau_{\text{max}} ],\omega \! \in\![ -\omega_{\text{max}},\omega_{\text{max}} ]   \right \}\!,
\end{align*}
where the maximum normalized delay $\tau_{\text{max}}$ and the maximum normalized Doppler $\omega_{\text{max}}$ are given by $\tau_{\text{max}} \ge  \frac{4d_{\text{max}}B}{c}$ and $\omega_{\text{max}} \ge  \frac{4v_{\text{max}}f_{\text{cr}}N}{cB}$, $c$ is the speed of light, $B$ is the bandwidth and $f_{\text{cr}}$ is the carrier frequency. For brevity, we simplify the notation of $\bm \Gamma({\tau_{\text{max}},\omega_{\text{max}}})$ to $\bm \Gamma$ in the following discussion. In addition, to simplify the analysis of computational complexity, we assume that the order of magnitude of $\tau_{\text{max}} \omega_{\text{max}}$ can be expressed in terms of $K^2$ throughout the paper.
\subsection{Traditional High-Complexity Exhaustive Search Method}
For sequences with traditional spike-like AAF, in the absence of any prior knowledge about the delay-Doppler ZoO, the identification of a peak corresponding to a certain target delay and Doppler requires an exhaustive search across the entire delay-Doppler plane. This holds true regardless of whether a periodic or aperiodic AAF with a symmetric or asymmetric receive reference sequence is utilized. If a “point-by-point" search is performed, the total (real-time processing) complexity is $\mathcal{O}(N^3)$. With the help of FFT, a “line-by-line" calculation scheme can reduce the complexity \cite{Toole2010}. Specifically, any “line" in the delay-Doppler plane is a certain convolution that can be computed with a complexity of $\mathcal{O}(N \text{log}N)$. Therefore, the overall exhaustive search complexity of all $N$ “lines" is reduced to $\mathcal{O}(N^2 \text{log}N)$. When the ZoO is known, the real-time processing complexity of the ``point-by-point" search method is $\mathcal{O}(K^2N)$, whereby the complexity of the ``line-by-line" strategy is $\mathcal{O}(KN \log N)$.
\subsection{Flag Sequence with peak-curtain AAF}
Before introducing the Flag method \cite {Fish2013} for low-complexity delay-Doppler estimation, we first present the definition of a Flag sequence, which is essential for the Flag method. Fig.~\ref{AF_ideal_HWS}(a) provides an illustration of the ideal \emph{peak-curtain} AAF for the Flag method. In this work, sequences possessing such approximate \emph{peak-curtain} AAF shape are referred to as Flag sequences. Specifically, for a Flag sequence with a normalized energy of 1, the \emph{peak} of its AAF with an amplitude of 1 at the origin is utilized to capture the delay-Doppler shift. Unlike the spike-like AAF for the exhaustive search method, the ideal AAF of the Flag sequence required by the Flag method has a \emph{curtain} part with an amplitude of 0.5 along a line $\bm l$ that passes through the origin in the delay-Doppler plane (with the exception of the origin itself). Thus, the ideal \emph{peak-curtain} AAF for Flag sequence can be expressed as:
\begin{equation}
	\label{FMAF}
	A_{\text{Flag}}(\tau,\omega)=\begin{cases}
	    1,\quad & (\tau,\omega)=(0,0); \\
        0.5,\quad & (\tau,\omega) \in \bm l \setminus (0,0);\\
        0,\quad & \text{otherwise},
	\end{cases}
\end{equation}
where “$\setminus$" denotes exception. 
\subsection{Almost-Linear-Complexity Flag Method}
\label{FMCC}
\begin{figure}[htb]
  \centering
  \begin{minipage}[b]{0.492\linewidth}
    \centering
    \includegraphics[trim=0.32cm 0cm 0.5cm 0.5cm, clip = true, width=\linewidth]{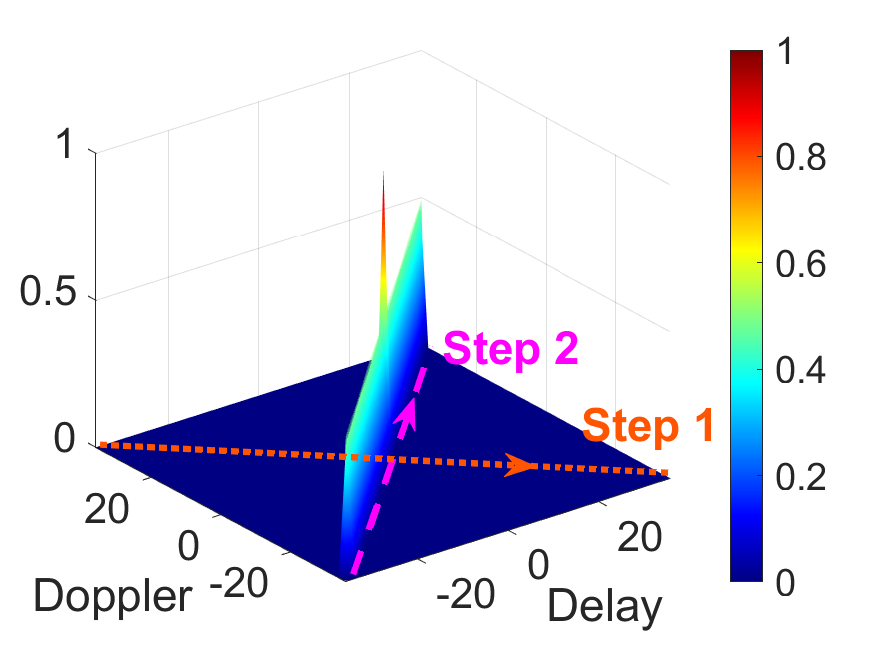}
    \footnotesize{(a)}
  \end{minipage}
  \hfill
  \begin{minipage}[b]{0.492\linewidth}
    \centering
    \includegraphics[trim=0.32cm 0cm 0.5cm 0.5cm, clip = true, width=\linewidth]{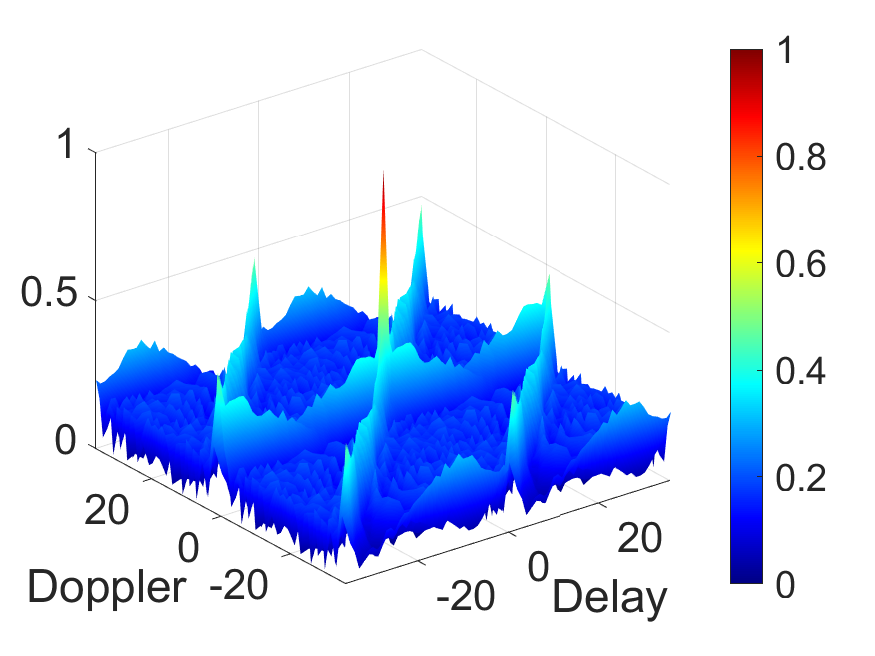}
    \footnotesize{(b)}
  \end{minipage}
  \begin{minipage}[b]{0.492\linewidth}
    \centering
    \includegraphics[trim=0.32cm 0cm 0.5cm 0.5cm, clip = true, width=\linewidth]{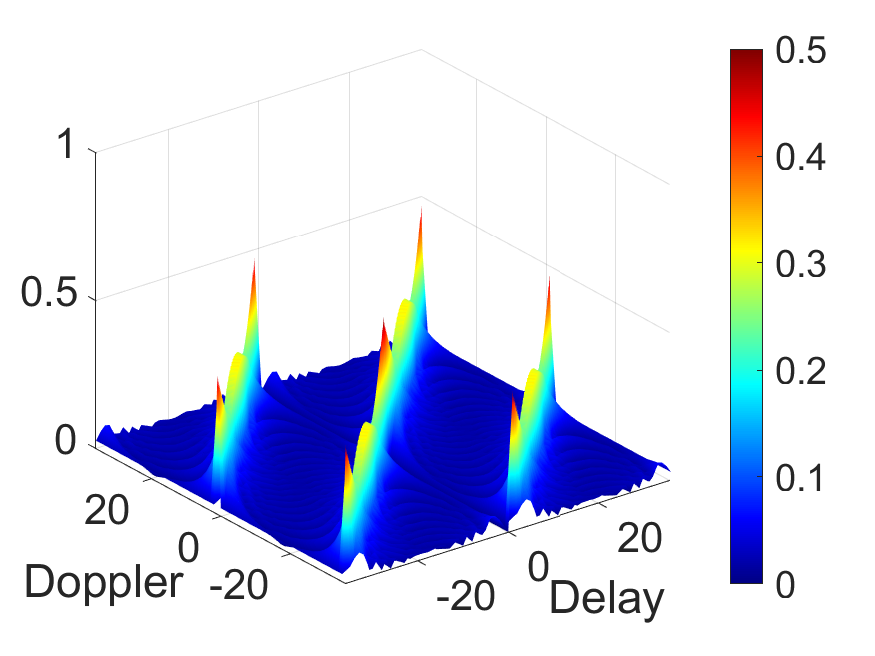}
    \footnotesize{(c)}
  \end{minipage}
  \hfill
  \begin{minipage}[b]{0.492\linewidth}
    \centering
    \includegraphics[trim=0.32cm 0cm 0.5cm 0.5cm, clip = true, width=\linewidth]{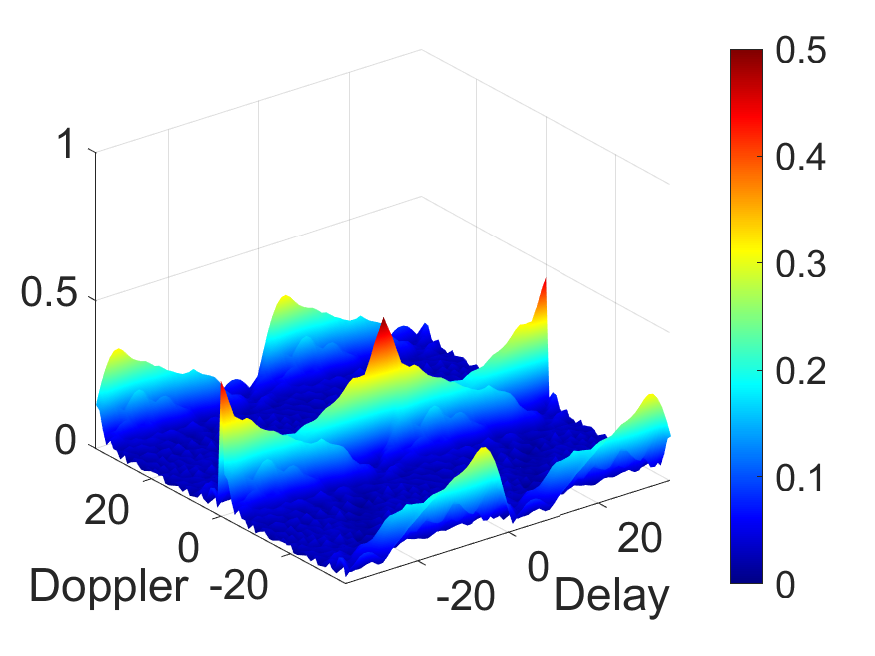}
    \footnotesize{(d)}
  \end{minipage}
  \caption{(a) Ideal auto-AF (AAF) of Flag sequences and illustration of the Flag method in 2-step search; (b) Periodic AAF of a Heisenberg-Weil sequence (HWS) with a length of 37; (c) Periodic AAF of the Heisenberg sequence component (\emph{curtain}) of the HWS shown in Fig.~\ref{AF_ideal_HWS}(b); (d) Periodic AAF of the Weil sequence component (\emph{peak}) of the HWS shown in Fig.~\ref{AF_ideal_HWS}(b).}
  \label{AF_ideal_HWS}
\end{figure}

The Flag method \cite {Fish2013} is a 2-step search method to determine the delay and Doppler values of AAF peaks by using the Flag sequence. When an echo $\bm s^{e}$ with a delay-Doppler of $(\tau_0,\omega_0)$ of the transmit sequence $\bm s$ is received, the output $A_{\bm s^{e},\bm r}$ of receive filter with reference sequence $\bm r$ approximately yields a noise-affected shifted version of the AAF $A_{\bm s,\bm r}$, where all points in $A_{\bm s,\bm r}$ are shifted by $\tau_0$ along the delay axis and $\omega_0$ along the Doppler axis \cite{Skolnik2008}. If $A_{\bm s,\bm r}$ is a \emph{peak-curtain} AAF similar to $A_{\text{Flag}}$, to identify the maximum point or the \emph{peak} of $A_{\bm s^{e},\bm r}$, the Flag method only requires two searches with almost linear complexity \cite {Fish2013}. Based on the delay and Doppler of the target, following the definition in (\ref{FMAF}), the \emph{curtain} is moved from line $\bm l$ to an unknown line $\vec{\bm{l}}$. Note that the direction of the \emph{curtain} is pre-designed and known, so the direction of shifted line $\vec{\bm{l}}$ is also known. Thus, \textbf{Step 1} of the Flag method involves finding the \emph{curtain} in a pre-defined direction transversal to the \emph{curtain}. Then, \textbf{Step 2} is carried out along the direction of the \emph{curtain} to locate the \emph{peak}. Specifically, the Flag method can be described as follows:
\begin{description}
    \item[Step 1:]Select a line $\hat{\bm l}$ transversal to line $\bm l$ (such as the orange dotted arrow in Fig.~\ref{AF_ideal_HWS}(a)). Compute $A_{\bm s^{e},\bm r}(\tau,\omega)$ with $(\tau,\omega) \in \hat{\bm l}$ by FFT, constituting a linear search. Identify $(\tau',\omega')$ such that $A_{\bm s^{e},\bm r}(\tau',\omega')$ exceeds a threshold $\eth$, i.e., finding \emph{curtain at} the shifted line $\vec{\bm l}$ of $\bm l$.
    \item[Step 2:]Compute $A_{\bm s^{e},\bm r}(\tau,\omega)$ with $(\tau,\omega) \in \vec{\bm l}$ by FFT, corresponding to a linear search in the direction parallel to the \emph{curtain} direction (indicated by the pink dashed arrow in Fig.~\ref{AF_ideal_HWS}(a)). Find $(\tau'',\omega'')$ such that $A_{\bm s^{e},\bm r}(\tau'',\omega'')$ exceeds the sum of $\eth$ and the average value of $A_{\bm s^{e},\bm r}(\tau,\omega)$ over $(\tau,\omega) \in \vec{\bm l} \setminus (\tau'',\omega'')$, i.e., finding the \emph{peak} at $(\tau_0,\omega_0)$.
\end{description}

Similarly, in applications with multiple targets or paths, the first search locates multiple \emph{curtains}, whereas the second search locates all the \emph{peaks} of different targets by searching all these \emph{curtains}. Therefore, the Flag method can provide a fast delay-Doppler estimation with an almost linear (real-time processing) computational complexity of $\mathcal{O}(N\text{log}N)$. 
\subsection{Basic Architecture and the Existing Design of the Flag Sequence}
A straightforward approach for constructing a \emph{peak-curtain} AAF is to combine two sequences that are almost mutually orthogonal, with one sequence having a \emph{peak} shaped AAF and the other having a \emph{curtain} shaped AAF \cite{Fish2013}. In this paper, a sequence with a \emph{peak} shaped AAF is referred to as a Peak sequence. The design objectives of a Peak sequence should consider the orthogonality with the Curtain sequence. A sequence with an approximate \emph{curtain} shaped AAF is called a Curtain sequence, and its ideal AAF can be expressed as follows:
\begin{equation}
	\label{curtainAF}
	A_{\text{Curtain}}(\tau,\omega)=\begin{cases}
	    1,\quad & (\tau,\omega) \in \bm l;\\
        0,\quad & \text{otherwise}.
	\end{cases}
\end{equation} 

Then, we can represent a Flag sequence as
\begin{equation}
	\bm f = (\bm c+\bm p)/\sqrt{2},
\end{equation}
where $\bm c$ denotes a Curtain sequence and $\bm p$ represents a Peak sequence. Note that the receive reference sequences may differ from the transmit sequences. 

In \cite{Fish2013}, the authors studied periodic \emph{peak-curtain} AAFs for sequence lengths limited to prime numbers. The idea is to use a Weil sequence as the Peak sequence, and a Heisenberg sequence as the Curtain sequence, in order to construct a novel family of Flag sequences called HWS. Fig.~\ref{AF_ideal_HWS}(b) shows the periodic AAF for an HWS of length 37, whose explicit expression is provided in \cite[Section IV-C2]{Fish2013}. One can observe that its AAF suffers from a poor \emph{curtain}. This is due to the poor \emph{curtain} performance\footnote{Note that the term `poor \emph{curtain} performance' here does not refer to the short curtains on either side in Fig.~\ref{AF_ideal_HWS}(c), as they can be excluded from the ZoO through reasonable design based on practical applications. The main issue is that the long \emph{curtain} passing through the origin is not uniform.} of the AAF in Fig.~\ref{AF_ideal_HWS}(c). For Fig.~\ref{AF_ideal_HWS}(d), the AAFs of Weil sequences do not possess distinct \emph{peak} shapes. Although some Heisenberg sequences can achieve the ideal \emph{curtain} AAF, a comprehensive study is missing in the literature. 

In light of these observations, we study novel designs for Flag sequences that consider periodic/aperiodic AF with both symmetric reference sequences and asymmetric reference sequences, respectively, aiming to attain improved \emph{peak-curtain} shapes in the ZoO.

\section{Proposed Curtain Sequence Sets}
\label{Section3}
In this section, we unveil the novel principles of generating Curtain sequence sets of arbitrary length, having ideal \emph{curtain} AAFs and zero/near-zero CAFs in the ZoO for periodic and aperiodic cases, respectively. The proposed design of Curtain sequence sets serves as the cornerstone for developing new Flag sequence sets. In this section, we only consider symmetric receive reference sequence, as it is sufficient for generating Curtain sequences with ideal AAFs.
\subsection{Proposed Curtain Sequences}

We first consider single Curtain sequences. To attain the ideal \emph{curtain} AAF as shown in (\ref{curtainAF}), we realize Curtain sequences of arbitrary length through careful selection of parameters for discrete chirp sequences. A discrete chirp sequence $\bm c_{\xi,q}$ of length $N$ can be expressed as:
\begin{equation}
\label{chirp}
	c_{\xi,q}[n] \!=\! \frac{1}{\sqrt{N}} \text{exp} \!\left (\frac{jn(\xi n\!+\!q)\pi}{N} \right )\!,n=0,1,\cdots\!,N-1, 
\end{equation}
where $\xi, q \in [1-N,2-N, \cdots, N-1]$ are the chirp rate and a phase shift index, respectively.

\subsubsection{Proposed Curtain Sequences for Periodic AAFs}
Based on the nature of discrete chirp sequences, we first present a theorem on Curtain sequences targeting the periodic \emph{curtain} AAF.

\begin{theorem}[Periodic AAF]
\label{T1}
Any discrete chirp sequence $\bm c_{\xi,q}$ that satisfies $[\xi N-q]_2 = 0$ is an ideal Curtain sequence in terms of its periodic AAF. The \emph{curtain} of its AAF is located on the line $\omega=\xi\tau$ in the delay-Doppler ZoO. This holds for any sequence length $N$ and ZoO $\bm \Gamma $ that satisfy $\left|\xi \right|\tau_{\text{max}}+\omega_{\text{max}}<N$.
\end{theorem}
\begin{IEEEproof}
See Appendix~\ref{T1proof}.
\end{IEEEproof}

\begin{remark}
\label{remark1}
According to Theorem~\ref{T1}, the Heisenberg sequences in \cite{Fish2013} can be classified into the following three categories; note that they are only of prime lengths.
    \begin{itemize}
        \item The first category of Heisenberg sequences is essentially prime length Delta functions $ \bm \delta_{u}$, where $u \in [0,1,\cdots, N-1]$. $\delta_{u}[n] = 1$ if $n = u$, otherwise $\delta_{u}[n] = 0$. Such Delta functions exhibit ideal \emph{curtain} AAFs whose directions align with the Doppler axis. However, these Heisenberg sequences' large PAPR make them unsuitable for practical applications. The PAPR of a sequence $\bm s$ can be expressed as:
\begin{equation}
\text{PAPR} \left (\bm s \right ) = N\frac{\max_{n \in {0,1,\cdots,N-1}} { \left |s [n] \right |}^2}{{\left \| \bm s \right \|}^2}.
\label{PAPRC}
\end{equation}
\item The second category of Heisenberg sequences pertains to discrete chirp sequences with prime lengths $N$ and $\xi \in [0, 1, \cdots, N-1]$ but do not satisfy Theorem~\ref{T1}. Thus, their AAFs are not ideal \emph{curtains}.
        \item The third category of Heisenberg sequences is discrete chirp sequences that satisfy Theorem~\ref{T1}. These Heisenberg sequences are special cases of Theorem~\ref{T1} with prime lengths $N$ and $\xi \in [0, 1, \cdots, N-1]$.
    \end{itemize}
\end{remark}

\subsubsection{Proposed Curtain Sequences for Aperiodic AAFs}
Moreover, we consider ideal aperiodic \emph{curtain} AAF. For any sequences $\bm s$ and $\bm r$, both of length $N$ and with a constant amplitude of $1/\sqrt{N}$, we have
    \begin{align}
    \label{L1}
        {\left |A_{\bm s,\bm r}(\tau, \omega) \right |}^2 \le \frac{\left (N- |\tau|\right )^2}{N^2}, \quad \forall \,\, |\tau|, |\omega| \in [0, N-1].
    \end{align}
Therefore, it is impossible to achieve ideal \emph{curtain} aperiodic AAF with constant amplitude transmit sequences. To address this, we propose to utilize extended discrete chirp sequences as the receive reference sequences. The corresponding extended receive reference sequence $\overleftrightarrow{\bm c_{\xi,q}}$ with a length of $L = N + 2\tau_{\text{ext}}$ for a transmit sequence $\bm c_{\xi,q}$ is defined as:
\begin{align}
\label{extentedchirp}
    & \overleftrightarrow{c_{\xi,q}} [n] = \frac{1}{\sqrt{N}} \text{exp} \left (\frac{jn(\xi n+q)\pi}{N} \right ) ,\nonumber \\& n=-\tau_{\text{ext}}, 1-\tau_{\text{ext}} \cdots,N-1+\tau_{\text{ext}}.
\end{align}
Such an approach utilizes the cyclic structure of the discrete chirp sequences by a minor modification to the receive reference sequence. This effectively breaks the limitation imposed by (\ref{L1}) without requiring the periodic retransmission of the transmit sequences. Throughout the paper, we use $\overleftrightarrow{\cdot}$ to represent the extended receive reference sequence. To facilitate the computation of AFs, we define $\overline{\bm c_{\xi,q}}$ to denote a zero-padded version of the sequence $\bm c$. The length of this zero-padded sequence is the same as that of the extended receive reference sequence, i.e. we have $\overline{\bm c_{\xi,q}} = [\bm 0_{1 \times \tau_{\text{ext}}}, \bm c_{\xi,q}^{\text{T}}, \bm 0_{1 \times \tau_{\text{ext}}}]^{\text{T}}$. 

\begin{theorem} [Aperiodic AAF]
\label{T2}
For any discrete chirp sequence $\bm c_{\xi,q}$, its zero-padded version $\overline{\bm c_{\xi,q}}$ and its corresponding extended receive reference sequence $\overleftrightarrow{\bm c_{\xi,q}}$ with $\tau_{\text{ext}} < \left \lfloor \frac{N}{\left|\xi \right|}\right \rfloor$ have ideal \emph{curtain} aperiodic AAF in any ZoO $\bm \Gamma$ that satisfies $\tau_{\text{max}} \le \tau_{\text{ext}}$ and $\left|\xi \right|\tau_{\text{max}}+\omega_{\text{max}}<N$. The \emph{curtain} is located on the line $\omega=\xi\tau$ in the ZoO.
\end{theorem}

We omit the proof of Theorem~\ref{T2} as it is similar to the proof of Theorem~\ref{T1}. It is worth mentioning that the extension of the receive reference sequence introduces a slight LPG \cite{Qazi2012}, which can be expressed as $\text{LPG}_{\text{ext}} = 10\text{log}_{10} \frac{\left |{\overleftrightarrow{\bm c_{\xi,q}}}^{\dagger}{\overline{\bm c_{\xi,q}}} \right |^2}{{\left \| \bm c_{\xi,q} \right \|}^2 {\left \| \overleftrightarrow{\bm c_{\xi,q}} \right \|}^2 } = 10\text{log}_{10} \frac{N}{L}.$ In this paper, to differentiate from the asymmetric receive reference sequence, we still refer to the extended receive reference sequence as an extended symmetric receive reference sequence.
\subsection{Proposed Curtain Sequence Sets}
In order to generate Curtain sequence sets for low mutual interference, we present new approaches based on the properties of the CAF between different proposed Curtain sequences. For brevity, we define two discrete chirp sequences as $\bm a = \bm c_{\xi_a,q_a}$ and $\bm b= \bm c_{\xi_b,q_b}$, corresponding to unspecified values of $\xi_a,q_a,\xi_b$ and $q_b$ in this subsection. 

\subsubsection{Proposed Curtain Sequence Sets for Periodic AFs}
We first present the constructions of Curtain sequence sets with near-zero periodic CAFs in the ZoO.

\begin{corollary} [Low Periodic CAF]
\label{C1}
For any two Curtain sequences $\bm a$ and $\bm b$ of length $N$ obtained from Theorem~\ref{T1}, let us assume that the following condition is satisfied: $[\xi_a N-q_a]_2 = [\xi_b N-q_b]_2 = 0$. In this case, their periodic CAF is constant and equal to $1/\sqrt{N}$ for any $(\tau,\omega)$, provided that $\left |\xi_a-\xi_b \right |$ and $N$ are relatively prime.
\end{corollary}
\begin{IEEEproof}
See Appendix~\ref{C1proof}.
\end{IEEEproof}

Leveraging Corollary~\ref{C1}, Curtain sequence sets that exhibit near-zero periodic CAFs in the ZoO can be generated. The maximum number of sequences in such a set equals the cardinality of the largest subset obtained from $[1-N, 2-N, \cdots, N-1]$ with the absolute difference between any two elements within this subset is relatively prime to $N$. 

Furthermore, we present the principle of constructing Curtain sequence sets with zero periodic CAFs in the ZoO. 

\begin{corollary}[Zero Periodic CAF]
\label{C3}
Given any two of our proposed Curtain sequences $\bm a$ and $\bm b$ of length $N$ that satisfy $\xi_a = \xi_b = \xi$, $[q_a-q_b]_2 = 0$ and $\left||q_a|-|q_b| \right| = 2d$, one has zero periodic CAF in $\bm \Gamma$ that satisfies $\left|\xi \right|\tau_{\text{max}}+\omega_{\text{max}}<d$. Conversely, for a given $\bm \Gamma$, the maximum number of sequences in this Curtain sequence set is $\min \left \{ \left \lfloor \frac{N}{|\xi|(\tau_{\text{max}}+1)}  \right \rfloor, \left \lfloor \frac{N}{\omega_{\text{max}}+1}  \right \rfloor \right \}$.
\end{corollary}
The proof of Corollary~\ref{C3} is similar to that of Corollary~\ref{C1} and hence omitted here. Based on Corollary~\ref{C3}, Curtain sequence sets with zero CAFs in the ZoO can be generated.

\subsubsection{Proposed Curtain Sequence Sets for Aperiodic AFs}
Then, we present the constructions of Curtain sequence sets with near-zero aperiodic CAFs in the ZoO.

\begin{corollary}[Low Aperiodic CAF]
\label{C2}
Let $\bm a$ and $\bm b$ be two Curtain sequences of length $N$ obtained from Theorem~\ref{T2}, and their corresponding extended receive reference sequences $\overleftrightarrow{\bm a}$ and $\overleftrightarrow{\bm b}$, respectively, with $\tau_{\text{ext}}\le \left \lfloor \frac{N}{2 \text{max}\left \{ \xi_a,\xi_b\right \} }\right \rfloor$. If $\left |\xi_a-\xi_b \right |$ is relatively prime to $N$, then the aperiodic CAF between $\overline{\bm a}$ and $\overleftrightarrow{\bm b}$ or that of $\overline{\bm b}$ and $\overleftrightarrow{\bm a}$ will be constant and equal to $1/\sqrt{N}$ for any $(\tau,\omega)$ that satisfies $\tau \le \tau_{\text{ext}}$.
\end{corollary}

\begin{corollary}[Zero Aperiodic CAF]
\label{C4}
Let us consider any two of our proposed Curtain sequences $\bm a$ and $\bm b$ of length $N$ and their extended receive reference sequences $\overleftrightarrow{\bm a}$ and $\overleftrightarrow{\bm b}$ with $\tau_{\text{ext}}$. When $\xi_a = \xi_b = \xi$, $[q_a-q_b]_2 = 0$ and $\left||q_a|-|q_b| \right| = 2d$, the aperiodic CAF of $\overline{\bm a}$ and $\overleftrightarrow{\bm b}$ or that of $\overline{\bm b}$ and $\overleftrightarrow{\bm a}$ is zero in any $\bm \Gamma$ that satisfies $\left|\xi \right|\tau_{\text{max}}+\omega_{\text{max}}<d$ and $\tau_{\text{max}} \le \tau_{\text{ext}}$. Conversely, for a given $\bm \Gamma$ with $\tau_{\text{max}} \le \tau_{\text{ext}}$, the maximum number of sequences in this Curtain sequence set is $\min \left \{ \left \lfloor \frac{N}{|\xi|(\tau_{\text{max}}+1)}  \right \rfloor, \left \lfloor \frac{N}{\omega_{\text{max}}+1}  \right \rfloor \right \}$.
\end{corollary}

Again, we omit the proof of Corollary~\ref{C2} and Corollary~\ref{C4} as they are similar to the proof of Corollary~\ref{C1}.


\section{Proposed Design of Flag Sequence Sets}
In this section, based on our proposed Curtain sequence sets, we design Flag sequence sets. We will first focus on the problem formulation for the asymmetric transmit sequences and receive reference sequences and show the symmetric transmit sequences and receive reference sequences case without extra effort. By reducing the WImSL of the Flag sequences, the sidelobes outside the \emph{peak-curtain} and the fluctuations of the \emph{curtain} are minimized, thus preventing false alarms and the obscuration of weak targets. Consequently, the proposed Flag sequences can be applied across various scenarios where spike-like AAF sequences are applicable.
\subsection{Definitions}
For both the periodic and the aperiodic cases with a ZoO $\bm \Gamma$, we use unified notations to represent the set of Flag sequences as $\bm F ^{s} =\left \{  \bm f_{m}^{s}\right \} _{m=1}^{M}$ and the set of receive reference sequences as $\bm F ^{r} =\left \{  \bm f_{m}^{r}\right \} _{m=1}^{M}$, where
\begin{subequations}
    \label{tsrs}
    \begin{align}
        & f_{m}^{s} [n] = \left (c_{m}^{s} [n] + p_{m}^{s} [n] \right )/ \sqrt{2}, \\
        & f_{m}^{r} [n] = \left (c_{m}^{r} [n] + p_{m}^{r} [n] \right )/ \sqrt{2}. 
    \end{align}
\end{subequations}
Moreover, we have two cases:
\begin{enumerate}
    \item \emph{Periodic AF Case of (\ref{tsrs}):} The sequence lengths of $\bm f_{m}^{s}$ and $\bm f_{m}^{r}$ are $L = N$, with $n = 0,1,\cdots, N-1$. We define $\bm c_{m}^{s} = \bm c_{m}^{r} = \bm c_{\xi_m,q_m}$ based on a proposed periodic AF case Curtain sequence set $\left \{  \bm c_{\xi_m,q_m}\right \} _{m=1}^{M}$. The Peak sequence $\bm p_{m}^{s}$ and the reference Peak sequence $\bm p_{m}^{r}$ can be expressed as
        \begin{align*}
            & \bm p_{m}^{s} = [p_{m}^{s}[0],p_{m}^{s}[1],\cdots,p_{m}^{s}[N-1]]^{\text{T}}, \\
            & \bm p_{m}^{r} = [p_{m}^{r}[0],p_{m}^{r}[1],\cdots,p_{m}^{r}[N-1]]^{\text{T}}.
        \end{align*}
    \item \emph{Aperiodic AF Case of (\ref{tsrs}):} The sequence lengths of $\bm f_{m}^{s}$ and $\bm f_{m}^{r}$ are $L = N+2 \tau_{\text{max}}$, $n = -\tau_{\text{max}},1-\tau_{\text{max}},\cdots, N-1+\tau_{\text{max}}$. Let us consider a proposed aperiodic AF case Curtain sequence set $\left \{  \bm c_{\xi_m,q_m}\right \} _{m=1}^{M}$ and the corresponding extended receive reference sequence set $\left \{  \overleftrightarrow{\bm c_{\xi_m,q_m}}\right \} _{m=1}^{M}$ with $\tau_{\text{ext}}=\tau_{\text{max}}$. Then, we define $\bm c_{m}^{s} = [\bm 0_{1 \times \tau_{\text{max}}},{\bm c_{m}}^{\text{T}},\bm 0_{1 \times \tau_{\text{max}}}]^{\text{T}}$ and $\bm c_{m}^{r} = \overleftrightarrow{\bm c_{m}}$. The Peak sequence $\bm p_{m}^{s}$ and the reference Peak sequence $\bm p_{m}^{r}$ can be written as
        \begin{align*}
            & \bm p_{m}^{s} = [\bm 0_{1 \times \tau_{\text{max}}},p_{m}^{s}[0],\cdots,p_{m}^{s}[N-1], \bm 0_{1 \times \tau_{\text{max}}}]^{\text{T}}, \\
            & \bm p_{m}^{r} = [\bm 0_{1 \times \tau_{\text{max}}},p_{m}^{r}[0],\cdots,p_{m}^{r}[N-1],\bm 0_{1 \times \tau_{\text{max}}}]^{\text{T}}.
        \end{align*}
\end{enumerate}
The unified representation in (\ref{tsrs}) is introduced for brevity in our subsequent discussions. In the rest of this section, we will not delve into the differentiation of periodic/aperiodic AF cases. Additionally, we define the Peak sequence set as $\bm P ^{s} =\left \{  \bm p_{m}^{s}\right \} _{m=1}^{M}$ and the reference Peak sequence set as $\bm P ^{r} =\left \{  \bm p_{m}^{r}\right \} _{m=1}^{M}$.
\subsection{Objective Function}
For any transmit sequence $\bm f_{m_1}^{s}$ and any receive reference sequence $\bm f_{m_2}^{r}$, our objective is to ensure that $\bm A_{\bm f_{m_1}^{s},\bm f_{m_2}^{r}}$ approximates the ideal \emph{peak-curtain} shape within $\bm \Gamma$ when $m_1 = m_2$. Moreover, we aim to minimize all the sidelobes of $\bm A_{\bm f_{m_1}^{s},\bm f_{m_2}^{r}}$ in $\bm \Gamma$ when $m_1 \neq m_2$. To achieve these objectives, we consider a unified metric, WImSL, as the objective function extended from WISL \cite{Song2016,Arlery2016s, Cui2017}, which is defined as
\begin{align}
    G  \left (\bm P ^{s} , \bm P ^{r}  \right) & = \alpha \sum_{m=1}^{M} S \left (\bm f_{m}^{s},\bm f_{m}^{r} \right ) \nonumber \\
    +  &(1-\alpha)\sum_{m_1=1}^{M} \sum_{\substack{m_2=1\\m_2 \neq m_1}}^{M} S \left (\bm f_{m_1}^{s},\bm f_{m_2}^{r} \right ),
	\label{obj}
\end{align}
where $G(\bm{P}^s, \bm{P}^r)$ represents the objective function that integrates the masked sidelobe levels in the ZoO across all AAFs and CAFs of the transmit and receive sequence set. $\alpha \in [0,1]$ is a weight factor between the AAF and CAF parts. $S(\bm{f}_{m_1}^s, \bm{f}_{m_2}^r)$ denotes the WImSL for the AF between transmit sequence $\bm{f}_{m_1}^s$ and receive reference sequence $\bm{f}_{m_3}^s$, which can be expressed as
\begin{align}
     S &\left (\bm f_{m_1}^{s},\bm f_{m_2}^{r} \right ) = \sum_{(\tau,\omega) \in \bm \Gamma }  \big | W_{m_1,m_2}(\tau,\omega) {\bm p_{m_1}^{s}}^{\dagger}  \bm U_{\tau,\omega} \bm p_{m_2}^{r} \nonumber \\&+ \overline{W}_{m_1,m_2}(\tau,\omega) \left ({\bm p_{m_1}^{s}}^{\dagger}  \bm U_{\tau,\omega} \bm c_{m_2}^{r}+{\bm c_{m_1}^{s}}^{\dagger}  \bm U_{\tau,\omega} \bm p_{m_2}^{r}\right ) \nonumber\\
    &+ \widetilde{W}_{m_1,m_2} {\bm c_{m_1}^{s}}^{\dagger}  \bm U_{\tau,\omega} \bm c_{m_2}^{r}  \big | ^2,
	\label{CWISL}
\end{align}
where $\bm U_{\tau, \omega}=\bm J_{\tau}\text{Diag}(\bm h(\omega))$. Let $\varrho \ge 1$ be the weight for origins of the AFs between Curtain sequences and the Peak sequences, we define
\begin{equation}
    W_{m_1,m_2}(\tau,\omega) = \begin{cases}
	    0,\quad & (\tau,\omega)  = (0,0) \text{ and } m_1 = m_2;\\
        1,\quad & \text{otherwise},
	\end{cases}
\end{equation}
\begin{equation}
    \overline{W}_{m_1,m_2}(\tau,\omega) = \begin{cases}
	    \varrho,\quad & (\tau,\omega)  = (0,0) \text{ and } m_1 = m_2;\\
        1,\quad & \text{otherwise},
	\end{cases}
\end{equation}
and 
\begin{equation}
    \widetilde{W}_{m_1,m_2} = \begin{cases}
	    0,\quad &  m_1 = m_2;\\
        1,\quad & \text{otherwise}.
	\end{cases}
\end{equation}
The WImSL in (\ref{obj}) includes all sidelobes of AAFs and CAFs in the ZoO except the ideal \emph{peak-curtain} in AAFs.
\subsection{Constraints of Interest}
\label{Section3C}
\label{conssubs}
Given that the Curtain sequences remain fixed during the optimization process, we consider the following constraints on the Peak sequences:
\begin{enumerate}[]
\item At the transmitting side, to control the PAPR fluctuation of the transmit sequences, we impose the constraint that $\bm p_{m}^{s}$ have constant amplitude in (\ref{PA}b), i.e.,
\begin{align}
    &\left |p_{m}^{s}[n] \right | = \frac{1}{ \sqrt{N}},\nonumber\\ &n=0,1, \cdots,N-1,\; m=1,2, \cdots,M.
    \label{cons1}
\end{align}

\item At the receiving side, to control the energy of the receive reference sequences, in (\ref{PA}c), we constrain
\begin{align}
    &{\left \| \bm p_{m}^{r} \right \|} ^2 = 1,\nonumber\\ &n=0,1, \cdots,N-1,\; m=1,2, \cdots,M.
	\label{energyC}
\end{align}
\item The use of asymmetric receive reference sequences introduces greater design degree-of-freedom, allowing for further sidelobe suppression, but it also results in LPG. To control the LPG, we constrain the peak magnitude $A_{\bm p_{m}^{s},\bm p_{m}^{r}}(0,0)$, which can be written as
\begin{equation}
    \left | { \bm p_{m}^{s}}^{\dagger}  \bm p_{m}^{r} \right | \gtrsim \epsilon ,\quad m = 1,2,\cdots,M,
	\label{peakC}
\end{equation}
where $\epsilon$ denotes the predefined Peak magnitude of the Peak sequences. This constraint is incorporated as a penalty function, represented by the second term in (\ref{PA}a). The corresponding LPG is given by:
\begin{equation}
\label{LPG}
\begin{split}
    \text{LPG} \left( \bm f_{m}^{s},\bm f_{m}^{r} \right) = 10\text{log}_{10} \frac{|{\bm f_{m}^{r}}^{\dagger}{\bm f_{m}^{s}}|^2}{{\left \| \bm f_{m}^{s} \right \|}^2 {\left \|  \bm f_{m}^{r} \right \|}^2 }.
\end{split}
\end{equation}
\end{enumerate}

Note the orthogonality between $\bm{p}_m^s$ and $\bm{c}_m^r$ as well as $\bm{p}_m^r$ and $\bm{c}_m^s$, i.e., ${\bm{p}_m^s}^\dagger \bm{c}_m^r$ and ${\bm{c}_m^r}^\dagger \bm{p}_m^s$ are incorporated into the objective function with weight $\varrho$. Meanwhile, the similarity between $\bm{p}_m^s$ and $\bm{c}_m^s$ as well as $\bm{p}_m^r$ and $\bm{c}_m^r$ is ensured by their definitions and the constraint (\ref{peakC}). Thus, by minimizing (\ref{obj}) under the above constraints, $\bm{p}_m^s$ and $\bm{c}_m^s$ as well as $\bm{p}_m^r$ and $\bm{c}_m^r$ become approximately orthogonal, i.e.
\begin{equation}
    \Delta=\max_{m}\left\{\max\left({\bm{p}_m^s}^\dagger \bm{c}_m^s, {\bm{c}_m^r}^\dagger \bm{p}_m^r \right)\right \} \approx 0.
    \label{assert}
\end{equation}
We verified (\ref{assert}) in Section~\ref{NRsub1}. On this basis, constraint (\ref{cons1}) at the transmitting side is equivalent to limiting the PAPR of the transmit sequence $\bm{f}_{m}^{s}$ to approximately $2$ (which corresponds to approximately 3.01~dB). On the receiving side, constraint (\ref{energyC}) essentially requires that the energy of $\bm{f}_{m}^{r}$ is almost $1$. Also, from constraint (\ref{peakC}), it can be inferred that $\text{LPG} \left( \bm f_{m}^{s},\bm f_{m}^{r} \right) \approx \text{LPG} \left( \bm p_{m}^{s},\bm p_{m}^{r} \right)/2 = 10 \log_{10}{\epsilon}$. Thus, $\epsilon$ can be selected based on the acceptable LPG between $\bm f_{m}^{s}$ and $\bm f_{m}^{r}$. In other words, the above constraints on Peak sequences effectively regulate the Flag sequences.
\subsection{Optimization Problems}
Note that the constraint (\ref{peakC}) can be added to the objective function as a penalty function. Thus, the optimization problem for asymmetric transmit sequences and receive reference sequences can be formulated as follows:
\begin{subequations}
	\label{PA}
    \begin{alignat}{4}
	\min\limits_{\bm P ^{s} , \bm P ^{r}}\quad& \beta G  \left (\bm P ^{s} , \bm P ^{r} \right) + (1- \beta) \sum_{m=1}^{M} {\left | {\bm p_{m}^{s}}^{\dagger} \bm p_{m}^{r} - \epsilon \right | }^2\\ 
 \mbox{s.t.}\quad & |p_{m}^{s} [n]| = 1/\sqrt{N},\\
                  & {\left \| \bm p_{m}^{r} \right \|} ^2 = 1,\\
                  &   n=0,1, \cdots,N-1,\; m=1,2, \cdots,M,
	\end{alignat}
\end{subequations}
where $\beta \in [0,1]$ is a weight factor between the WImSL and the penalty.

Similarly, for symmetric transmit sequences and receive reference sequences, we have $\bm P ^{s} = \bm P ^{r} = \bm P = \left \{  \bm p_{m}\right \} _{m=1}^{M}$, the problem can be simplified as:
\begin{subequations}
	\label{PB}
    \begin{alignat}{4}
	\min\limits_{\bm P}\quad&G  \left (\bm P \right)\\ 
 \mbox{s.t.}\quad & |\bm p_{m} [n]| = 1/\sqrt{N},\\
                  &   n=0,1, \cdots,N-1,\; m=1,2, \cdots,M.
	\end{alignat}
\end{subequations}

We have formulated two optimization problems (\ref{PA}) and (\ref{PB}) for asymmetric/symmetric transmit sequences and receive reference sequences, respectively. These two problems are challenging because the objective functions and constraint sets in these problems are non-convex. In the subsequent section, we will present the AP-MM algorithms to tackle these two problems.
\section{Proposed WImSL Minimization under Majorization-Minimization Framework}
In this section, we propose efficient AP-MM algorithms to solve the above non-convex optimization problems. We begin by introducing the AP-MM algorithm, which is based on the MM framework and utilizes power method-like (PML) iterations to efficiently solve the optimization problem (\ref{PA}). The proposed algorithm can be further applied to the optimization problem (\ref{PB}) with minor changes. Examples of designed sequences are given in \url{https://github.com/meng0071/Flag_sequence}. 
\subsection{Proposed AP-MM Algorithm for Asymmetric Transmit Sequences and Receive Reference Sequences}
In this subsection, we consider the design of Flag sequence sets for asymmetric transmit sequences and receive reference sequences. For convenience of representation and analysis, we first synthesize $\bm P ^{s}$ and $\bm C ^{s}$ into a variable vector $\bm x_1$, and synthesize $\bm P ^{r}$ and $\bm C ^{r}$ into another variable vector $\bm x_2$, denoted as follows:
\begin{subequations}
\label{x1x2}
\begin{align}
    & \bm x_1 = {\left [ {\bm p_{1}^{s}}^{\text{T}}, {\bm c_{1}^{s}}^{\text{T}}, {\bm p_{2}^{s}}^{\text{T}}, {\bm c_{2}^{s}}^{\text{T}}, \cdots, {\bm p_{M}^{s}}^{\text{T}}, {\bm c_{M}^{s}}^{\text{T}} \right ] }^{\text{T}}, \\
    & \bm x_2 = {\left [ {\bm p_{1}^{r}}^{\text{T}}, {\bm c_{1}^{r}}^{\text{T}}, {\bm p_{2}^{r}}^{\text{T}}, {\bm c_{2}^{r}}^{\text{T}}, \cdots, {\bm p_{M}^{r}}^{\text{T}}, {\bm c_{M}^{r}}^{\text{T}} \right ] }^{\text{T}}.
\end{align}
\end{subequations}
We define \mbox{${\bm \Psi _{\tau, \omega}}= {\begin{bmatrix}
    W_{m,m}(\tau,\omega) \bm U_{\tau, \omega}&  \overline{W}_{m,m}(\tau,\omega)\bm U_{\tau, \omega} \\ 
     \overline{W}_{m,m}(\tau,\omega)\bm U_{\tau, \omega}& \bm 0_{L\times L}
    \end{bmatrix}}$} and ${\bm \Upsilon _{\tau, \omega}}= {\begin{bmatrix}
    \bm U_{\tau, \omega}& \bm U_{\tau, \omega} \\ 
    \bm U_{\tau, \omega}& \bm U_{\tau, \omega}
    \end{bmatrix}}$. Then (\ref{obj}) can be rewritten as 
\begin{align}
    G  \left (\bm x_1 , \bm x_2  \right) = &\alpha \sum_{i=1}^{M} \sum_{(\tau,\omega) \in \bm \Gamma } \left | {\bm x_2}^{\dagger} \hat{\bm B}_{\tau,\omega}^{i} {\bm x_1}\right | ^2 \nonumber\\
    +  (1- & \alpha)\sum_{l=1}^{M} \sum_{\substack{k=1\\k \neq l}}^{M} \sum_{(\tau,\omega) \in \bm \Gamma } \left | {\bm x_2}^{\dagger} \bm B_{\tau,\omega}^{l,k} {\bm x_1}\right | ^2,
    \label{obj2}
\end{align}
where 
\begin{align*}
    & \hat{\bm B}_{\tau,\omega}^{i} = \begin{bmatrix} \begin{smallmatrix}
  \bm 0_{2(i-1)L \times 2(i-1)L}& \bm 0_{2(i-1)L \times 2L} & \bm 0_{2(i-1)L \times 2(M-i)L}\\
  \bm 0_{2L \times 2(i-1)L}& \bm \Psi {\tau, \omega} & \bm 0_{2L \times 2(M-i)L}\\
  \bm 0_{2(M-i)L \times 2(i-1)L}& \bm 0_{2(M-i)L \times 2L} &\bm 0_{2(M-i)L \times 2(M-i)L}
\end{smallmatrix} \end{bmatrix} ,\\
& \bm B_{\tau,\omega}^{l,k} = \begin{bmatrix} \begin{smallmatrix}
  \bm 0_{2(k-1)L \times 2(l-1)L}& \bm 0_{2(k-1)L \times 2L} & \bm 0_{2(k-1)L \times 2(M-l)L}\\
  \bm 0_{2L \times 2(l-1)L}& \bm \Upsilon {\tau, \omega} & \bm 0_{2L \times 2(M-l)L}\\
  \bm 0_{2(M-k)L \times 2(l-1)L}& \bm 0_{2(M-k)L \times 2L} &\bm 0_{2(M-k)L \times 2(M-l)L}
\end{smallmatrix} \end{bmatrix}.
\end{align*}

After ignoring the constant terms, the optimization problem (\ref{PA}) can be transformed into
\begin{subequations}
	\label{PA2}
    \begin{alignat}{2}
       \min\limits_{\bm P ^{s} , \bm P ^{r}}\quad& G  \left (\bm x_1 , \bm x_2  \right) +\\
       & \beta' \left( \sum_{i=1}^{M}\left | {\bm x_2}^{\dagger} \bm M^{i} {\bm x_1} \right | ^2 - 2 \epsilon \Re \left \{  {\bm x_2}^{\dagger} \bm V {\bm x_1} \right \}  \right)\\
 \mbox{s.t.}\quad & |\bm p_{m}^{s} [n]| = 1/\sqrt{N},\\
                  & {\left \| \bm p_{m}^{r} \right \|} ^2 = 1,\\
                  & n=0,1, \cdots,N-1,\; m=1,2, \cdots,M,
	\end{alignat}
\end{subequations}
where 
\begin{align*}
    & \bm M^{i} =\\
    &\begin{bmatrix} \begin{smallmatrix}
  \bm 0_{2(i-1)L \times 2(i-1)L}& \bm 0_{2(i-1)L \times L} & \bm 0_{2(i-1)L \times (2M-2i+1)L}\\
  \bm 0_{L \times 2(i-1)L}& \bm I_{L} & \bm 0_{L \times (2M-2i+1)L}\\
  \bm 0_{(2M-2i+1)L \times 2(i-1)L}& \bm 0_{(2M-2i+1)L \times L} &\bm 0_{(2M-2i+1)L \times (2M-2i+1)L}
\end{smallmatrix} \end{bmatrix} ,
\end{align*}
with $\bm V = \sum_{i=1}^{M} \bm M^{i}$ and $\beta'= (1-\beta)/\beta$.

Then, we define ${\bm X}={\bm x_1}{\bm x_2}^{\dagger}$. Based on the property that ${\bm x_1}^{\dagger} {\hat{\bm B}}_{\tau,\omega}^{i} {\bm x_2} = \text{Tr} ({\hat{\bm B}}_{\tau,\omega}^{i} \bm X)= {\text{vec}(\bm X^{\dagger})}^{\dagger} \text{vec}({\hat{\bm B}}_{\tau, \omega}^{i})$, problem (\ref{PA2}) is equivalent to
\begin{subequations}
	\label{PA3}
    \begin{alignat}{4}
	\min\limits_{\bm P ^{s} , \bm P ^{r}}\quad & {\text{vec} (\bm X^{\dagger})}^{\dagger}{\bm \Lambda}{\text {vec} (\bm X^{\dagger})} -2 \beta' \epsilon \Re \left \{ {\bm x_2}^{\dagger} \bm V {\bm x_1} \right \}   \\
	 \mbox{s.t.}\quad & |p_{m}^{s} [n]| = 1/\sqrt{N},\\
                  & {\left \| \bm p_{m}^{r} \right \|} ^2 = 1,\\
                  & n=0,1, \cdots,N-1,\; m=1,2, \cdots,M,
	\end{alignat}
\end{subequations}
where $\bm \Lambda = \alpha \bm \Lambda_1+ (1-\alpha)\bm \Lambda_2 + \beta' \bm \Lambda_3$, 
\begin{subequations}
    \begin{align}
	\bm \Lambda_1=&\sum_{i=1}^{M} \sum_{(\tau,\omega) \in \bm \Gamma } {\text{vec}({\hat{\bm B}}_{\tau, \omega}^{i})}{\text{vec}({\hat{\bm B}}_{\tau, \omega}^{i})}^{\dagger},\\ 
        \bm \Lambda_2=& \sum_{j=1}^{M} \sum_{\substack{k=1\\k \neq l}}^{M} \sum_{(\tau,\omega) \in \bm \Gamma } {\text{vec}(\bm  B_{\tau, \omega}^{l,k})}{{\text{vec}(\bm  B_{\tau, \omega}^{l,k})}}^{\dagger},\\
        \bm \Lambda_3=& \sum_{i=1}^{M} {\text{vec}(\bm  M^{i})}{{\text{vec}(\bm  M^{i})}}^{\dagger}.
	\end{align}
\end{subequations}

It is noteworthy that $\bm \Lambda$ is a Hermitian matrix. For such a non-convex optimization problem, the MM framework can iteratively solve it by utilizing a surrogate problem.


\begin{proposition}
\label{P1}
The optimization problem (\ref{PA3}) can be majorized by the following problem at the $t$th iteration with $\bm X^{(t)} = {\bm x_1}^{(t)}{{\bm x_2}^{(t)}}^{\dagger}$:
\begin{subequations}
	\label{PA4}
    \begin{alignat}{4}
	\min\limits_{\bm P ^{s} , \bm P ^{r}}\quad & 2 \Re \left \{ {\text{vec} (\bm X^{\dagger})}^{\dagger}{\left( \bm \Lambda - \lambda \bm I_{(2ML)^2} \right)}{\text {vec} ({\bm X^{(t)}}^{\dagger})} \right \}  \\ &-2 \beta' \epsilon \Re \left \{ {\bm x_2}^{\dagger} \bm V {\bm x_1} \right \}  \\
	 \mbox{s.t.}\quad & |p_{m}^{s} [n]| = 1/\sqrt{N},\\
                  & {\left \| \bm p_{m}^{r} \right \|} ^2 = 1,\\
                  & n=0,1, \cdots,N-1,\; m=1,2, \cdots,M,
	\end{alignat}
\end{subequations}
where $\lambda > \lambda_{\text{max}}(\bm \Lambda)$, $\lambda_{\text{max}}(\bm \Lambda)$ represents the largest eigenvalue of $\bm \Lambda$.
\end{proposition}

\begin{IEEEproof}
    The proof of Proposition~\ref{P1} and the derivation of $\lambda_{\text{max}}(\bm \Lambda)$ are provided in Appendix~\ref{Ppproof}.
\end{IEEEproof}

By substituting $\bm X = {\bm x_1}{\bm x_2}^{\dagger}$ and $\bm X^{(t)}={\bm x_1}^{(t)}{{\bm x_2}^{(t)}}^{\dagger}$ back into (\ref{PA4}), the problem can be rearranged as
\begin{subequations}
	\label{PA5}
    \begin{alignat}{4}
	\min\limits_{\bm P ^{s} , \bm P ^{r}}\quad & 2 \Re \left \{ {\bm x_{2}}^{\dagger} \left( \bm \Omega_{\bm x_{1}^{(t)},\bm x_{2}^{(t)}} - \lambda \bm x_{2}^{(t)} {\bm x_{1}^{(t)}}^{\dagger}  \right) \bm x_{1}  \right \}  \\ 
    &-2 \beta' \epsilon \Re \left \{ {\bm x_2}^{\dagger} \bm V {\bm x_1} \right \}  \\
	 \mbox{s.t.}\quad & |p_{m}^{s} [n]| = 1/\sqrt{N},\\
                  & {\left \| \bm p_{m}^{r} \right \|} ^2 = 1,\\
                  & n=0,1, \cdots,N-1,\; m=1,2, \cdots,M,
	\end{alignat}
\end{subequations}
where 
\begin{align}
     &\bm \Omega_{\bm x_{1}^{(t)},\bm x_{2}^{(t)}} 
     =\begin{bmatrix}
      \widetilde{{\bm R}}_{1} & {\bm R}_{1,2} & \cdots & {\bm R}_{1,M}\\
      {\bm R}_{2,1}& \ddots  &\ddots  & \vdots \\
      \vdots&  \ddots& \ddots & {\bm R}_{M-1,M}\\
      {\bm R}_{M,1}&\cdots  & {\bm R}_{M,M-1} &\widetilde{{\bm R}}_{M}
    \end{bmatrix}.
\end{align}
with
\begin{subequations}
    \begin{align}
     &\widetilde{{\bm R}}_{i}= 
      \begin{bmatrix}
      \bm Q_{A}^{i}& \bm Q_{B}^{i} \\
      \bm Q_{B}^{i}& \bm 0_{L \times L}
    \end{bmatrix},  \quad {\bm R}_{i,l} = \begin{bmatrix}
      \bm Q_{C}^{i,l}& \bm Q_{C}^{i,l};\\
      \bm Q_{C}^{i,l}& \bm Q_{C}^{i,l};
    \end{bmatrix}, \\
    &\bm Q_{A}^{i} = \alpha \sum_{(\tau,\omega) \in \bm \Gamma}  W_{i,i}(\tau,\omega) {{\bm x_1}^{(t)}}^{\dagger} {{\hat{\bm B}}_{\tau,\omega}^{i}}{}^{\dagger} {\bm x_2}^{(t)} \bm U_{\tau,\omega} \label{usefftb} \\& \quad \quad \quad+ \beta' {{\bm x_1}^{(t)}}^{\dagger} {\bm M ^{i} }^{\dagger} {\bm x_2}^{(t)} \bm I_{L},\\
        &\bm Q_{B}^{i} = \alpha \sum_{(\tau,\omega) \in \bm \Gamma} {{\overline{W}_{m,m}(\tau,\omega)\bm x_1}^{(t)}}^{\dagger} {{\hat{\bm B}}_{\tau,\omega}^{i}}{}^{\dagger} {\bm x_2}^{(t)} \bm U_{\tau,\omega},\label{usefftd}\\ 
        &\bm Q_{C}^{i,l} = (1-\alpha) \sum_{(\tau,\omega) \in \bm \Gamma} {{\bm x_1}^{(t)}}^{\dagger} {\bm B_{\tau,\omega}^{i,l}}^{\dagger} {\bm x_2}^{(t)} \bm U_{\tau,\omega}, \label{useffte}\\
    &\text{for } i = 1,2,\cdots, M, \text{ and } l = 1,2,\cdots, M. \nonumber
\end{align}
\end{subequations}

At the $t$th iteration, the idea of alternating minimization \cite{Wang2022b} can be adopted by fixing ${\bm x_1} = {\bm x_1}^{(t)}$ to solve ${\bm x_2}^{(t+1)}$. Thus, (\ref{PA5}) can be simplified to
\begin{subequations}
	\label{PA6}
    \begin{alignat}{4}
	\min\limits_{\bm P ^{r}}\quad &  \Re \left \{ {\bm x_{2}}^{\dagger} \bm \kappa^{(t)}  \right \}  \\
	 \mbox{s.t.}\quad & {\left \| \bm p_{m}^{r} \right \|} ^2 = 1,\\
                  &  m=1,2, \cdots,M,
	\end{alignat}
\end{subequations}
where 
\begin{equation}
    \bm \kappa^{(t)} = \left( \bm \Omega_{\bm x_{1}^{(t)},\bm x_{2}^{(t)}} - \lambda \bm x_{2}^{(t)} {\bm x_{1}^{(t)}}^{\dagger} - \beta' \epsilon \bm V \right) \bm x_{1}^{(t)}.
\end{equation}
Since $\bm p_{m}^{r}$ and $\bm c_{m}^{r}$ are independent, we can optimize $\left \{  \bm p_{m}^{r}\right \} _{m=1}^{M}$ only while keeping $\left \{  \bm c_{m}^{r}\right \} _{m=1}^{M}$ fixed, and decompose problem (\ref{PA6}) into $M$ sub-problems that can be computed in parallel, i.e.,
\begin{subequations}
	\label{PA7}
    \begin{alignat}{4}
	\min\limits_{\bm p_{m} ^{r}}\quad &  \Re \left \{ {\bm p_{m} ^{r}}^{\dagger} \bm \iota_{2m-1}^{(t)} \right \}  \\
	 \mbox{s.t.}\quad & {\left \| \bm p_{m}^{r} \right \|} ^2 = 1,\\
                  &  m=1,2, \cdots,M,
	\end{alignat}
\end{subequations}
where $\bm \iota_{2m-1}^{(t)}$ is an $L \times 1$ sub-vector of $\bm \kappa^{(t)}$ with
\begin{align*}
    \bm \kappa^{(t)}= {\left [ {\bm \iota_{1}^{(t)}}^{\text{T}}, {\bm \iota_{2}^{(t)}}^{\text{T}}, \cdots, {\bm \iota_{2M}^{(t)}}^{\text{T}} \right ] }^{\text{T}}.
\end{align*}
Then, for each sub-problem, the Lagrange multipliers method can be used to obtain the optimal solution as
\begin{align}
    \label{PA7s}
        &{\bm p_{m}^{r}}^{(t+1)} = - \frac{\bm \iota_{2m-1}^{(t)}}{{\left \| \bm \iota_{2m-1}^{(t)} \right \|}  }. 
\end{align}
Using (\ref{PA7s}), we can obtain $\bm x_{2}^{(t+1)}$ by updating $\left \{{\bm p_{m}^{r}}^{(t+1)} \right \} _{m=1}^{M}$ while keeping $\left \{  \bm c_{m}^{r}\right \} _{m=1}^{M}$ unchanged.

Next, with fixed $\bm x_{2}^{(t+1)}$ and $\left \{  \bm c_{m}^{s}\right \} _{m=1}^{M}$, we solve $\bm x_{1}^{(t+1)}$ and recast the problem into $M$ sub-problems of $\left \{{\bm p_{m}^{s}}^{(t+1)} \right \} _{m=1}^{M}$ below:
\begin{subequations}
	\label{PA8}
    \begin{alignat}{4}
	\min\limits_{\bm p_{m} ^{s}}\quad &  \Re \left \{ \bm \gamma_{2m-1}^{(t)} \bm p_{m} ^{s}  \right \}  \\
	 \mbox{s.t.}\quad & |p_{m}^{s} [n]| = 1/\sqrt{N},\\
                  & n=0,1, \cdots,N-1,\; m=1,2, \cdots,M,
	\end{alignat}
\end{subequations}
where
\begin{align}
    &{\left [ {\bm \gamma_{1}^{(t)}}^{\text{T}}, {\bm \gamma_{2}^{(t)}}^{\text{T}}, \cdots, {\bm \gamma_{2M}^{(t)}}^{\text{T}} \right ] }^{\text{T}} \nonumber\\
    = &{\bm x_{2}^{(t+1)}}^{\dagger} \left( \bm \Omega_{\bm x_{1}^{(t)},\bm x_{2}^{(t+1)}} - \lambda \bm x_{2}^{(t+1)} {\bm x_{1}^{(t)}}^{\dagger} - \beta' \epsilon \bm V \right) .
\end{align}
The problem (\ref{PA8}) with the constant amplitude constraint can be solved with a closed form \cite{Soltanalian2013}, given by
\begin{align}
    \label{PA8s}
        {\bm p_{m}^{s}}^{(t+1)} = - \frac{1}{\sqrt{N}}\text{exp} \left( j \text{arg} \left( \bm \gamma_{2m-1} ^{(t)} \right) \right).
\end{align}

Finally, the optimization problems (\ref{PA7}) and (\ref{PA8}) can be accelerated using the two-point acceleration strategy \cite{Song2016, Cui2017}. Therefore, using $\text{OF}(\cdot)$ to denote the objective function of (\ref{PA}), the proposed AP-MM algorithm for asymmetric transmit sequences and receive reference sequences is summarized in Algorithm~1. 

\begin{algorithm}[tbp]
\caption{Proposed AP-MM algorithm for asymmetric transmit sequences and receive reference sequences} 
\hspace*{0.02in} {\bf Input:}
Initial Peak sequence set $\bm P^{s}$ and reference Peak sequences $\bm P^{r}$, ZoO $\bm \Gamma$, Curtain sequence set $\bm C$, anticipated $\epsilon$, weights $\varrho$, $\alpha$ and $\beta$;\\
\hspace*{0.02in} {\bf Output:} 
A Flag sequence set and its corresponding receive reference sequences;
\begin{algorithmic}[1]
\State Set $t=0$, initialize $\bm x_1^{(0)}$, and $\bm x_2^{(0)}$;
\Statex ${\textbf{repeat}}$
\State Based on $\bm x_{1}^{(t)}$ and $\bm x_{2}^{(t)}$, calculate ${\bm y_{m}^{a}}$ with (\ref{PA7s}),
\Statex $\bm y^a = {\left [ {\bm y_{1}^{a}}^{\text{T}}, {\bm c_{1}^{r}}^{\text{T}}, {\bm y_{2}^{a}}^{\text{T}}, {\bm c_{2}^{r}}^{\text{T}}, \cdots, {\bm y_{M}^{a}}^{\text{T}}, {\bm c_{M}^{r}}^{\text{T}} \right ] }^{\text{T}}$;
\State Based on $\bm x_{1}^{(t)}$ and $\bm y^{a}$, calculate ${\bm y_{m}^{b}}$ with (\ref{PA7s}),
\Statex $\bm y^b = {\left [ {\bm y_{1}^{b}}^{\text{T}}, {\bm c_{1}^{r}}^{\text{T}}, {\bm y_{2}^{b}}^{\text{T}}, {\bm c_{2}^{r}}^{\text{T}}, \cdots, {\bm y_{M}^{b}}^{\text{T}}, {\bm c_{M}^{r}}^{\text{T}} \right ] }^{\text{T}}$;
\State $\bm v_{m}^{a} = \bm y_{m}^{a} - {\bm p_{m}^{r}}^{(t)}$, $\bm v_{m}^{b} = \bm y_{m}^{b} - \bm y_{m}^{a}- \bm v_{m}^{a}$;
\State Compute the step length $\alpha_a = - \frac{\sum_{m=1}^{M}\left \| \bm v_{m}^{a} \right \|^2  }{\sum_{m=1}^{M}\left \| \bm v_{m}^{b} \right \|^2} $;
\State ${\bm p_{m}^{r}}^{(t+1)} = \frac{{\bm p_{m}^{r}}^{(t)}-2\alpha_a \bm v_{m}^{a}+ {\alpha_a}^{2}\bm v_{m}^{b}}{{\left \| {\bm p_{m}^{r}}^{(t)}-2\alpha_a \bm v_{m}^{a}+ {\alpha_a}^{2}\bm v_{m}^{b} \right \| } }$, 
\Statex update $\bm x_{2}^{(t+1)}$;
\State \quad \textbf{while} {$\text{OF}\left(\bm x_1^{(t)},\bm x_{2}^{(t+1)}\right)>\text{OF}\left(\bm x_1^{(t)},\bm x_{2}^{(t)}\right)$}
\Statex \quad \quad $\alpha_a = (\alpha_a -1)/2$, repeat Step 6;
\Statex \quad \textbf{end while}
\State Similar to Steps 2-7, update $\bm x_{1}^{(t+1)}$ based on (\ref{PA8s});
\State $t=t+1$;
\Statex ${\textbf{until}}$ convergence or $t=t_{\text{max}}$;
\end{algorithmic}
\end{algorithm}
\subsection{Proposed AP-MM Algorithm for Symmetric Transmit Sequences and Receive Reference Sequences}
In this subsection, we consider the design of Flag sequence sets for symmetric transmit sequences and receive reference sequences. For this purpose, (\ref{PA2}) to (\ref{PA5}) can be used again after removing the penalty term. Then, (\ref{PB}) can be rewritten as:
\begin{subequations}
	\label{PB1}
    \begin{alignat}{4}
	\min\limits_{\bm P}\quad & \Re \left \{ {\bm x}^{\dagger} \left( \bm \Omega_{\bm x^{(t)}} - \lambda \bm x^{(t)} {\bm x^{(t)}}^{\dagger}  \right) \bm x  \right \}  \\ 
	 \mbox{s.t.}\quad & |p_{m} [n]| = 1/\sqrt{N},\\
                  & n=0,1, \cdots,N-1,\; m=1,2, \cdots,M,
	\end{alignat}
\end{subequations}
where 
\begin{equation}
    \bm x = {\left [ {\bm p_{1}}^{\text{T}}, {\bm c_{1}}^{\text{T}}, {\bm p_{2}}^{\text{T}}, {\bm c_{2}}^{\text{T}}, \cdots, {\bm p_{M}}^{\text{T}}, {\bm c_{M}}^{\text{T}} \right ] }^{\text{T}}.
\end{equation}

Since $\left( \bm \Omega_{\bm x^{(t)}} - \lambda \bm x^{(t)} {\bm x^{(t)}}^{\dagger} \right)$ is not Hermitian, we first equivalently transform (\ref{PB1}) to
\begin{subequations}
	\label{PB2}
    \begin{alignat}{4}
	\min\limits_{\bm P}\quad &   {\bm x}^{\dagger} \left( \bm \Omega_{\bm x^{(t)}} + {\bm \Omega_{\bm x^{(t)}}}^{\dagger} - 2 \lambda \bm x^{(t)} {\bm x^{(t)}}^{\dagger}  \right) \bm x    \\ 
	 \mbox{s.t.}\quad & |p_{m} [n]| = 1/\sqrt{N},\\
                  & n=0,1, \cdots,N-1,\; m=1,2, \cdots,M.
	\end{alignat}
\end{subequations}

\begin{proposition}
\label{P2}
(\ref{PB2}) can be majorized at $\bm x^{(t)}$ as:
\begin{subequations}
	\label{PB3}
    \begin{alignat}{4}
	\min\limits_{\bm P}\,\, &  \Re \left \{ {\bm x}^{\dagger} \left( \bm \Omega_{\bm x^{(t)}} + {\bm \Omega_{\bm x^{(t)}}}^{\dagger} - 2 \lambda \bm x^{(t)} {\bm x^{(t)}}^{\dagger} - \widetilde{\lambda}  \bm I_{2ML}  \right) \bm x^{(t)} \right \}   \\ 
	 \mbox{s.t.}\,\, & |p_{m} [n]| = 1/\sqrt{N},\\
                  & n=0,1, \cdots,N-1,\; m=1,2, \cdots,M,
	\end{alignat}
\end{subequations}
where $\widetilde{\lambda} > \lambda_{\text{max}} \left( \bm \Omega_{\bm x^{(t)}} + {\bm \Omega_{\bm x^{(t)}}}^{\dagger} - 2 \lambda \bm x^{(t)} {\bm x^{(t)}}^{\dagger} \right)$.
\end{proposition}

\begin{IEEEproof}
    The proof of Proposition~\ref{P2} and the choice of $\widetilde{\lambda}$ are provided in Appendix~\ref{Ppproof}.
\end{IEEEproof}

\begin{algorithm}[htp]
\caption{Proposed AP-MM algorithms for symmetric transmit sequences and receive reference sequences} 
\hspace*{0.02in} {\bf Input:}
Initial Peak sequence set $\bm P$, ZoO $\bm \Gamma$, Curtain sequence set $\bm C$, weights $\varrho$ and $\alpha$;\\
\hspace*{0.02in} {\bf Output:} 
A Flag sequence set and its corresponding receive reference sequences;
\begin{algorithmic}[1]
\State Set $t=0$, initialize $\bm x_1^{(0)}$, and $\bm x_2^{(0)}$;
\Statex ${\textbf{repeat}}$
\State Based on $\bm x_{1}^{(t)}$, calculate ${\bm y_{m}^{a}}$ with (\ref{Pb8s}), 
\Statex $\bm y^a = {\left [ {\bm y_{1}^{a}}^{\text{T}}, {\bm c_{1}}^{\text{T}}, {\bm y_{2}^{a}}^{\text{T}}, {\bm c_{2}}^{\text{T}}, \cdots, {\bm y_{M}^{a}}^{\text{T}}, {\bm c_{M}}^{\text{T}} \right ] }^{\text{T}}$;
\State Based on ${\bm y_{m}^{a}}$, calculate ${\bm y_{m}^{b}}$ with (\ref{Pb8s}),
\Statex $\bm y^b = {\left [ {\bm y_{1}^{b}}^{\text{T}}, {\bm c_{1}}^{\text{T}}, {\bm y_{2}^{b}}^{\text{T}}, {\bm c_{2}}^{\text{T}}, \cdots, {\bm y_{M}^{b}}^{\text{T}}, {\bm c_{M}}^{\text{T}} \right ] }^{\text{T}}$;
\State $\bm v_{m}^{a} = \bm y_{m}^{a} - {\bm p_{m}}^{(t)}$, $\bm v_{m}^{b} = \bm y_{m}^{b} - \bm y_{m}^{a}- \bm v_{m}^{a}$;
\State Compute the step length $\alpha_s = - \frac{\sum_{m=1}^{M}\left \| \bm v_{m}^{a} \right \|^2  }{\sum_{m=1}^{M}\left \| \bm v_{m}^{b} \right \|^2} $;
\State ${\bm p_{m}^{r}}^{(t+1)} = \frac{1}{\sqrt{N}}\text{exp}\left( j \text{arg} \left({{\bm p_{m}}^{(t)}-2\alpha_s \bm v_{m}^{a}+ {\alpha_s}^{2}\bm v_{m}^{b}} \right)\right)$, 
\Statex update $\bm x^{(t+1)}$;
\State \quad \textbf{while} {$\text{OF}\left (\bm x^{(t+1)}\right)>\text{OF}\left(\bm x^{(t)}\right)$}
\Statex \quad \quad $\alpha_s = (\alpha_s -1)/2$, repeat Step 6;
\Statex \quad \textbf{end while}

\State $t=t+1$;
\Statex ${\textbf{until}}$ convergence or $t=t_{\text{max}}$;
\end{algorithmic}
\end{algorithm}

Similar to the asymmetric case, we can decompose problem (\ref{PB3}) into the following $M$ subproblems:
\begin{subequations}
	\label{PB4}
    \begin{alignat}{4}
	\min\limits_{\bm p_{m}}\quad &   \Re \left \{ {\bm x}^{\dagger} \bm \sigma_{2m-1}^{(t)} \right \}\\ 
	 \mbox{s.t.}\quad & |p_{m} [n]| = 1/\sqrt{N},\\
                  & n=0,1, \cdots,N-1,\; m=1,2, \cdots,M,
	\end{alignat}
\end{subequations}
where
\begin{align}
    &{\left [ {\bm \sigma_{1}^{(t)}}^{\text{T}}, {\bm \sigma_{2}^{(t)}}^{\text{T}}, \cdots, {\bm \sigma_{2M}^{(t)}}^{\text{T}} \right ] }^{\text{T}} \nonumber\\
    = & \left( \bm \Omega_{\bm x^{(t)}} + {\bm \Omega_{\bm x^{(t)}}}^{\dagger} - 2 \lambda \bm x^{(t)} {\bm x^{(t)}}^{\dagger} - \widetilde{\lambda}  \bm I_{2ML}  \right) \bm x^{(t)}.
\end{align}
Sub-problems (\ref{PB4}) can also be solved in closed form as \cite{Soltanalian2013}
\begin{align}
    \label{Pb8s}
        {\bm p_{m}}^{(t+1)} = -\frac{1}{\sqrt{N}} \text{exp} \left( j \text{arg} \left(\bm \sigma_{2m-1}^{(t)}\right) \right). 
\end{align}

Again, let $\text{OF}(\cdot)$ denote the objective function of (\ref{PB}). The proposed AP-MM algorithm for symmetric transmit sequences and receive reference sequences is summarized in Algorithm~2. 

\subsection{Offline Sequence Design Complexity Analysis}
In this subsection, we analyze the offline sequence design complexity of the proposed AP-MM algorithms.

The sequence design complexity of Algorithm~1 primarily arises from the operations in (\ref{usefftb}), (\ref{usefftd}), and (\ref{useffte}) for computing $\bm{\iota}_{2m-1}^{(t)}$ in (\ref{PA7s}) and $\bm{\gamma}_{2m-1}^{(t)}$ in (\ref{PA8s}). These computations can be efficiently performed using FFT, as described in \cite[Appendix~B]{Song2016} and \cite{Wang2021}. Specifically, in each iteration, the complexity of Steps 2-3 in Algorithm~1 is $\mathcal{O}(M^2KN\log N)$.

Additional vector operations and updates, such as step length calculations in Step 5 and sequence updates in Step 8, contribute lower-order complexities of $\mathcal{O}(MN)$. Consequently, the overall (one-time) offline sequence design complexity per iteration of Algorithm~1 is $\mathcal{O}(M^2KN\log N + MN)$, which is dominated by $\mathcal{O}(M^2KN\log N)$. Similarly, Algorithm~2 benefits from FFT, resulting in a (one-time) offline sequence design complexity of $\mathcal{O}(M^2KN\log N + MN)$ in each iteration.

It is important to note that this subsection addresses the offline sequence design complexity, which is distinct from the real-time processing complexity of delay-Doppler estimation discussed in Section~II.

\section{Numerical Results and Discussions}
In this section, we evaluate the performances of the proposed Flag sequences and compare them with the only existing Flag sequences namely HWS \cite{Fish2013}. We begin by comparing their AFs and associated metrics. Subsequently, we analyze their delay-Doppler detection and estimation performances in more practical scenarios. Note that we do not compare detection and estimation performances with traditional spike-like AAF sequences \cite{Cui2017, Arlery2016m} in this section because they have much higher estimation complexity (see Section~II). 

First, to further evaluate the performances of the AFs, we define the normalized WImSL (NWImSL) as
\begin{align}
        &\text{NWImSL}^{(t)} = 10\log_{10} \left ( \frac{G\left ( {\bm P^{s}}^{(t)},{\bm P^{r}}^{(t)} \right ) }{G\left ( {\bm P^{s}}^{(0)},{\bm P^{r}}^{(0)} \right )} \right ),
\end{align}
where $\text{WImSL}_{\text{ref}}$ refers to the WImSL of a benchmark sequence.
Also, we define a new metric called peak-to-max-masked-sidelobe ratio (PMmSR) to validate the superiority of our proposed Flag sequences over HWS in achieving \emph{peak-curtain}-like AAFs in the ZoO. Similar to the concept of the peak-to-max-sidelobe ratio (PMSR) for spike-like AAF sequences \cite{liu2022}, PMmSR represents the ratio of the AAF's \emph{peak} to the maximum sidelobe (including the fluctuation of the \emph{curtain}) within the ZoO, corresponding to the ideal \emph{peak-curtain}. The PMmSR is defined as:
    \begin{align}
    \label{PMSR}
        \text{PMmSR}_{\bm F^{s},\bm F^{r}} \!=\!  \underset{m}{\text{min}}\!\left\{\!\frac{\underset{(\tau,\omega) \in \bm \Gamma}{\text{max}}  \!\left |A_{\bm f^{s},\bm f^{r}}(\tau,\omega)\!-\!A_{\text{Flag}}(\tau,\omega) \right |  }{\left | A_{\bm f^{s},\bm f^{r}}(0,0) \right | }\!\right \}\!. 
    \end{align}

Without loss of generality, unless otherwise stated, we employ $\varrho=1$, $\alpha = 0.5$, and $\beta = 0.01$ in the AP-MM algorithms to equally consider AAFs and CAFs and impose strict constraint on the LPG. 
\subsection{AFs and WImSL Minimization}
\label{NRsub1}
In this subsection, we first provide a visual comparison between the AAFs of our proposed Flag sequence and the traditional HWS. We also present the AFs of our designed sequence sets for multi-user radar applications. We then demonstrate the evolution curves of the NWImSL with respect to the computational time by using our proposed AP-MM algorithms. Furthermore, we compare the NWImSL and the PMmSR of our proposed Flag sequences with those of HWS, and validate that the LPG of our proposed Flag sequences approaches the expected values. We also verify the assertion (\ref{assert}) through numerical examples.


\begin{figure*}[htbp]
    \centering
	\begin{minipage}[b]{0.65\columnwidth}
		\centering
            {\includegraphics[trim=2.7cm 0cm 2.5cm 1.4cm, clip = true, width=1\linewidth]{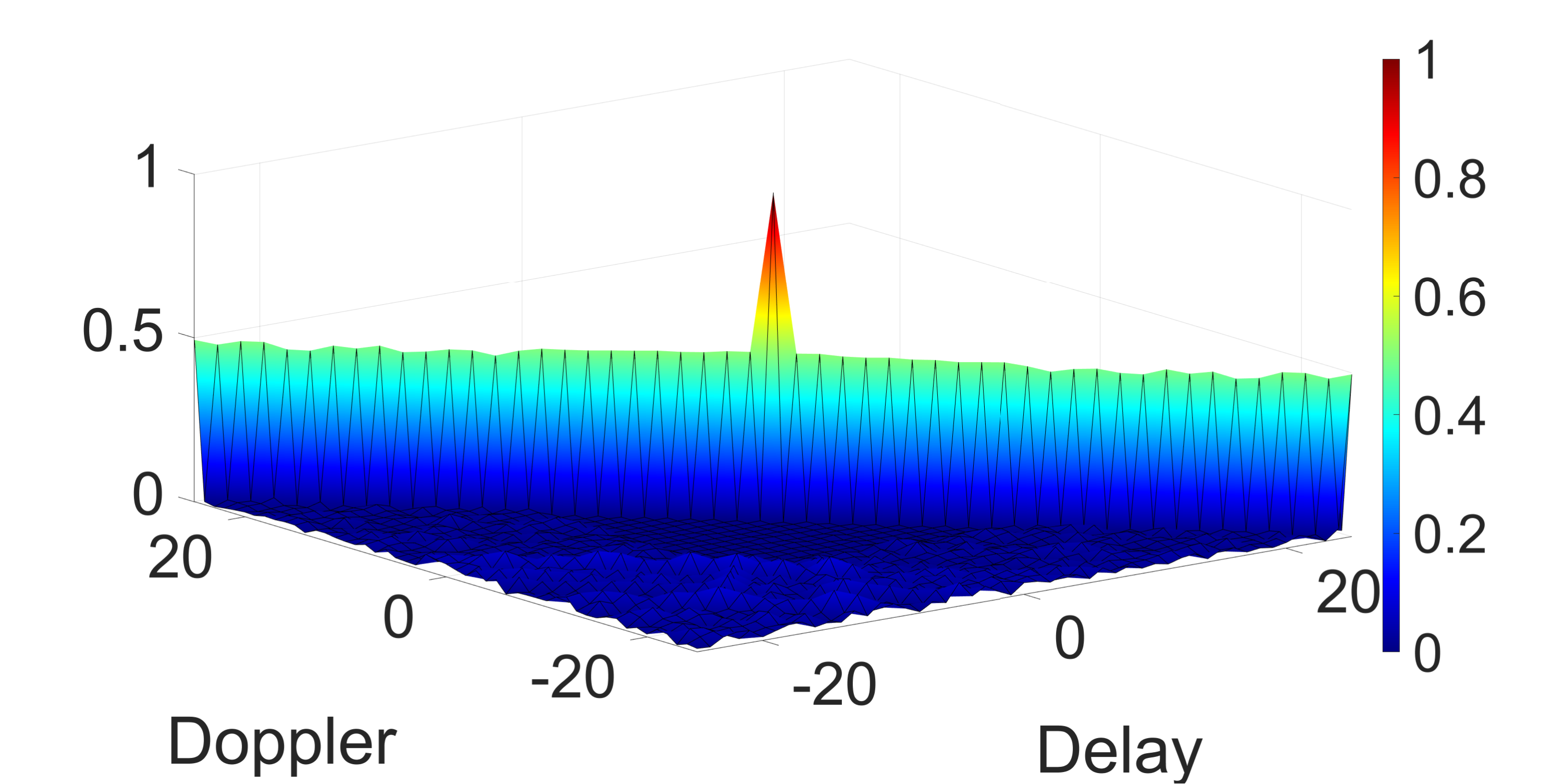}} \\ \footnotesize{(a)}\\
            {\includegraphics[trim=2.7cm 0cm 2.5cm 1.4cm, clip = true, width=1\linewidth]{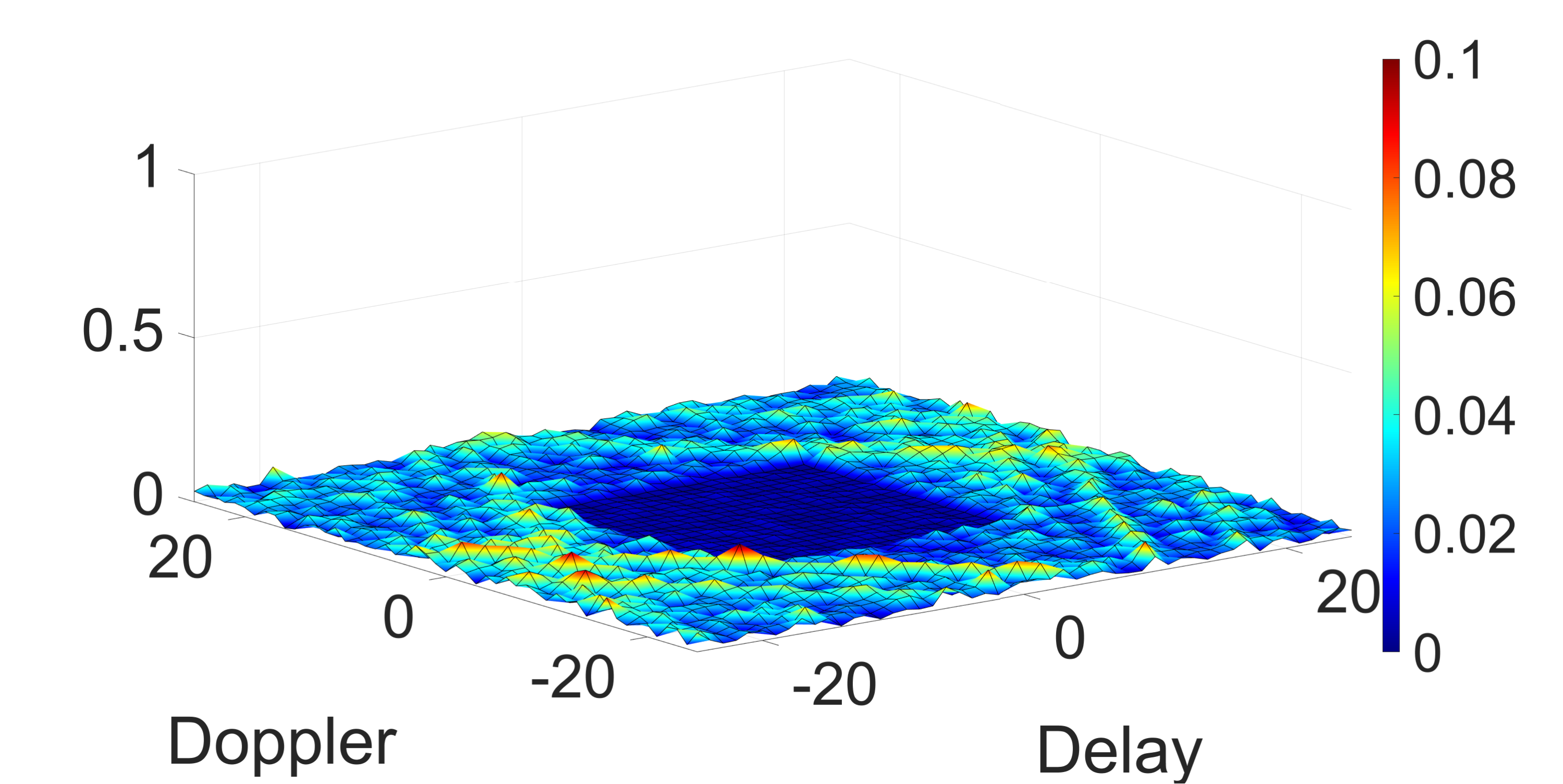}} \\ \footnotesize{(d)}\\
            {\includegraphics[trim=2.7cm 0cm 2.5cm 1.4cm, clip = true, width=1\linewidth]{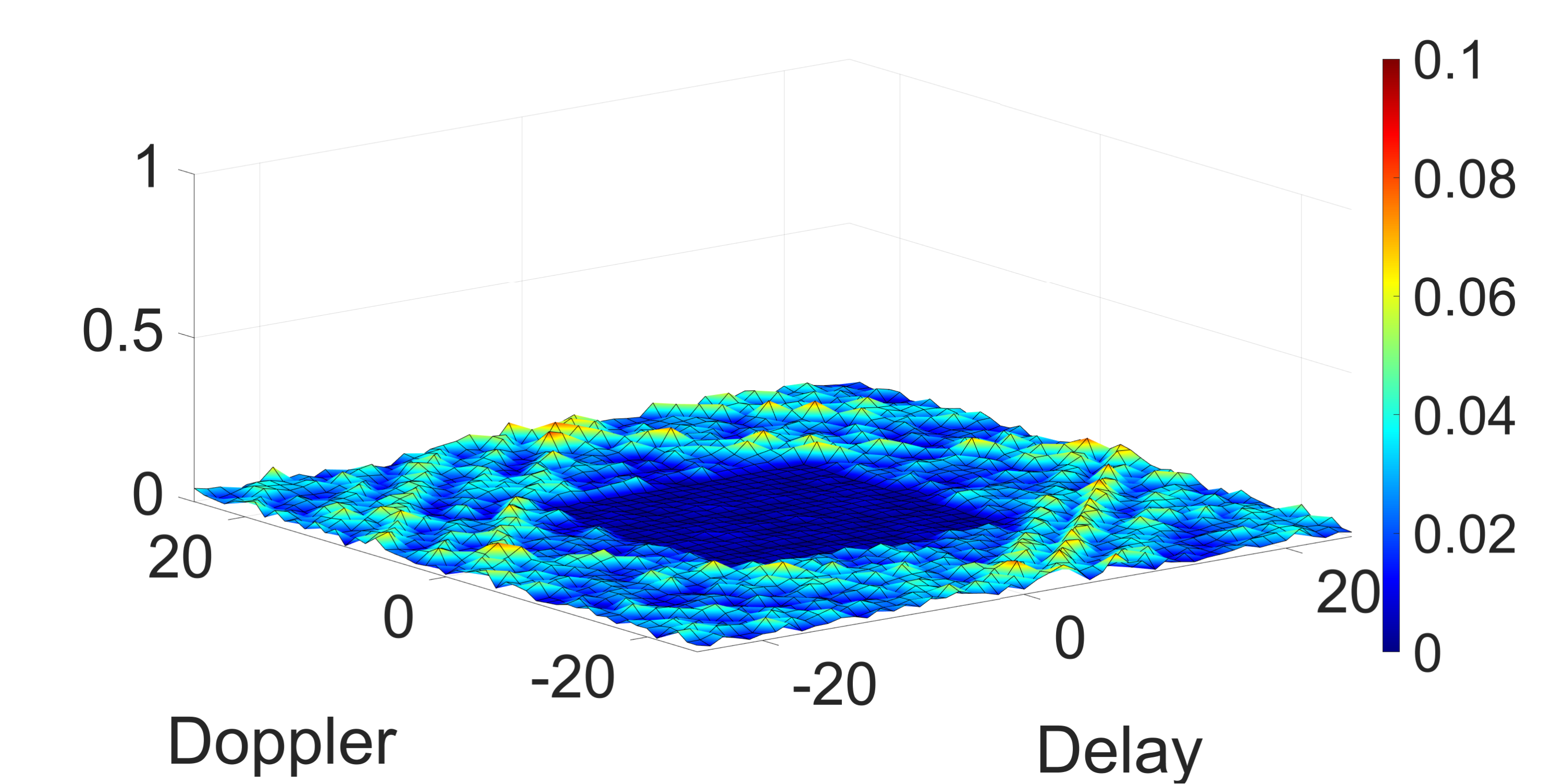}} \\ \footnotesize{(g)}\\
	\end{minipage}
	\begin{minipage}[b]{0.65\columnwidth}
		\centering
            {\includegraphics[trim=2.7cm 0cm 2.5cm 1.4cm, clip = true, width=1\linewidth]{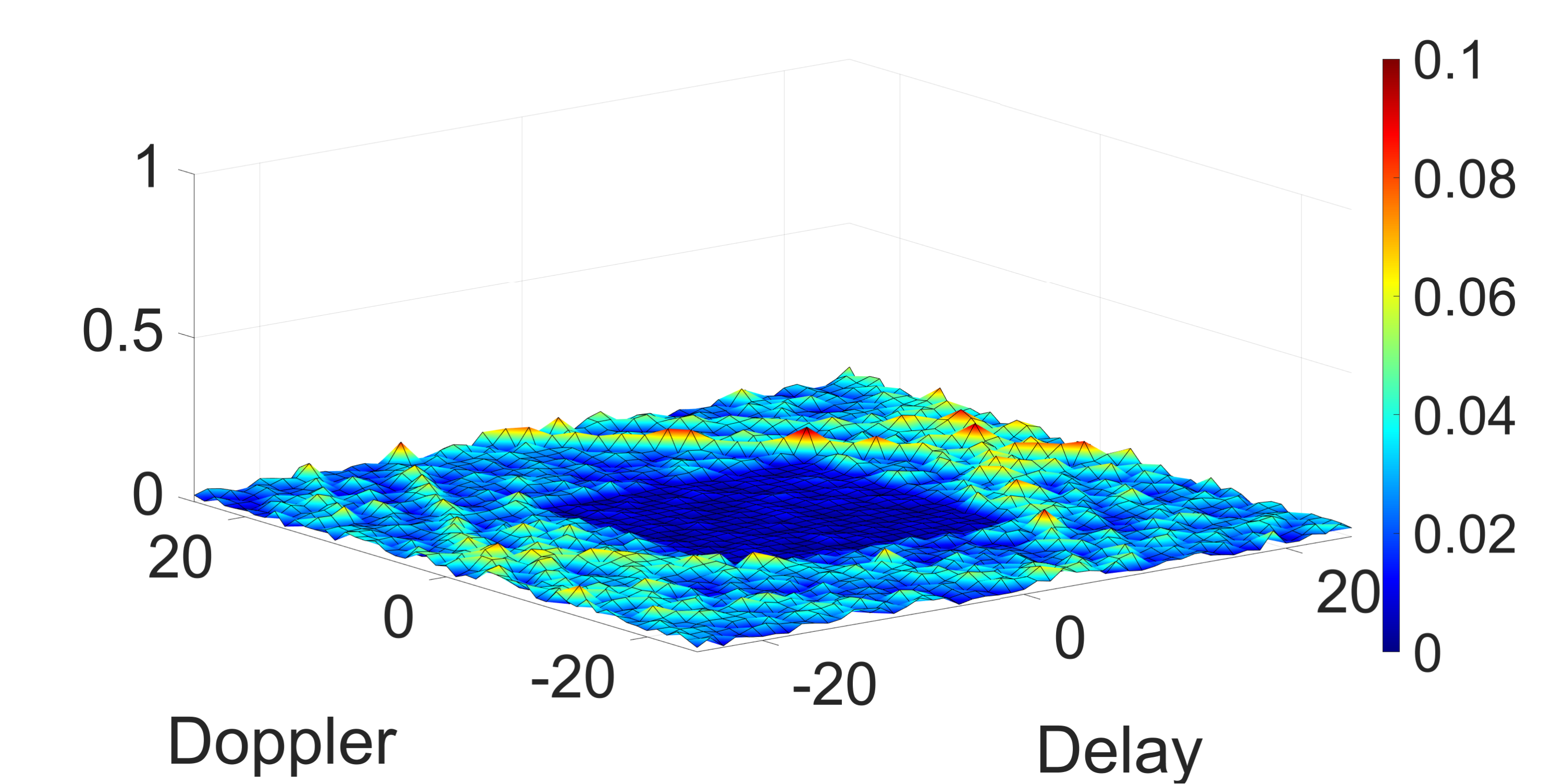}} \\ \footnotesize{(b)}\\
            {\includegraphics[trim=2.7cm 0cm 2.5cm 1.4cm, clip = true, width=1\linewidth]{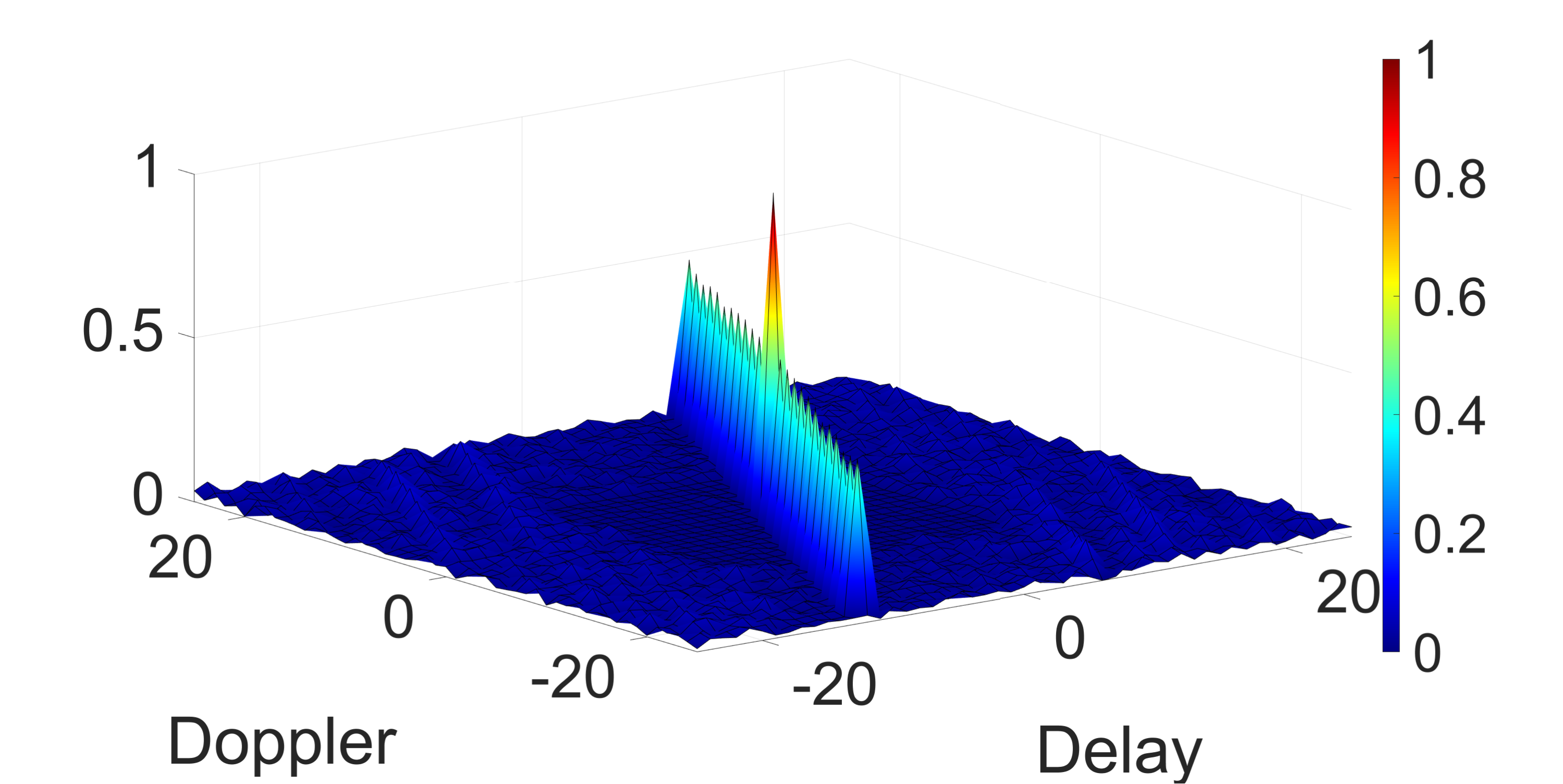}} \\ \footnotesize{(e)}\\
            {\includegraphics[trim=2.7cm 0cm 2.5cm 1.4cm, clip = true, width=1\linewidth]{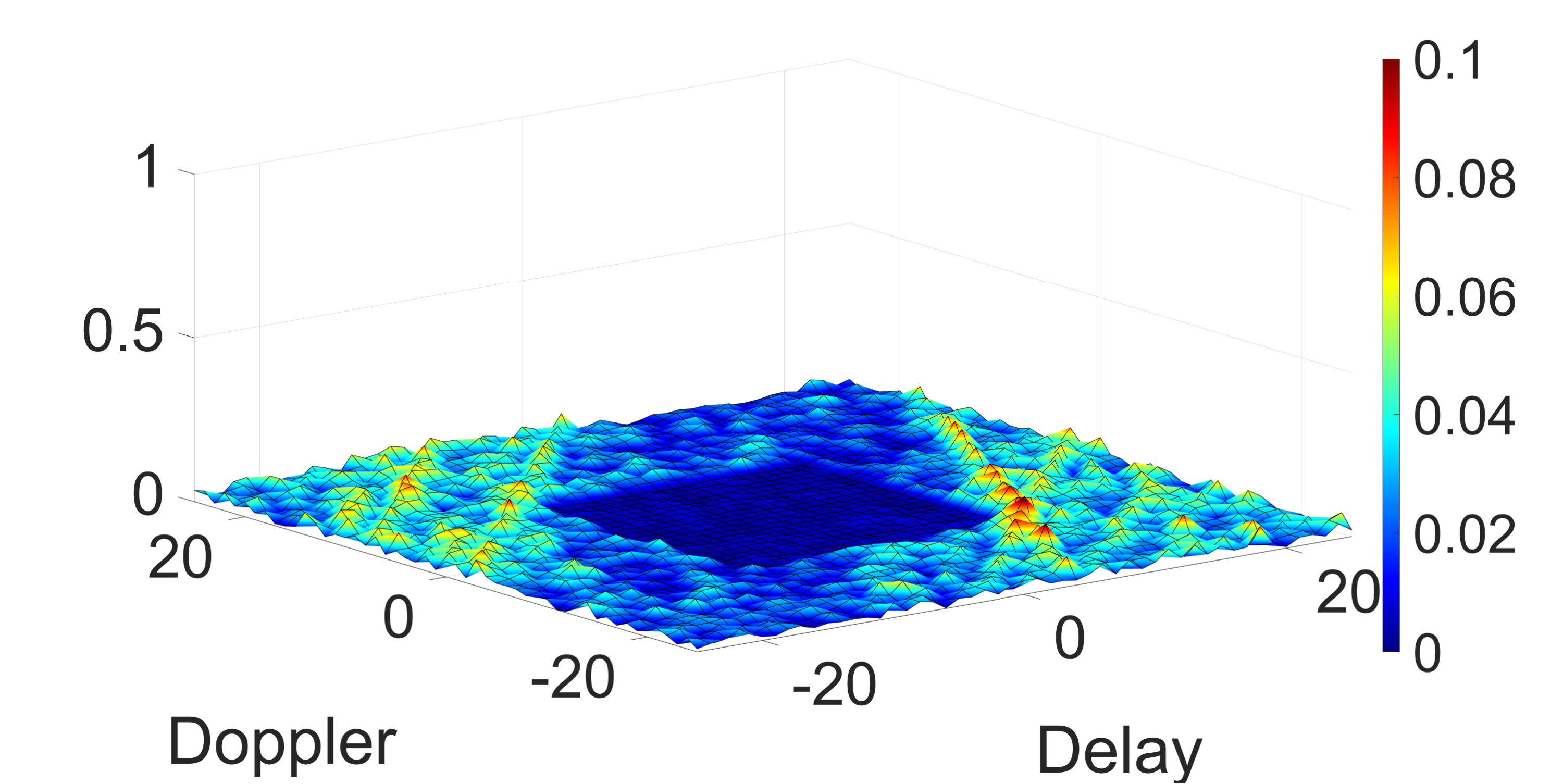}} \\ \footnotesize{(h)}\\
	\end{minipage}
 	\begin{minipage}[b]{0.65\columnwidth}
		\centering
            
        {\includegraphics[trim=2.7cm 0cm 2.5cm 1.4cm, clip = true, width=1\linewidth]{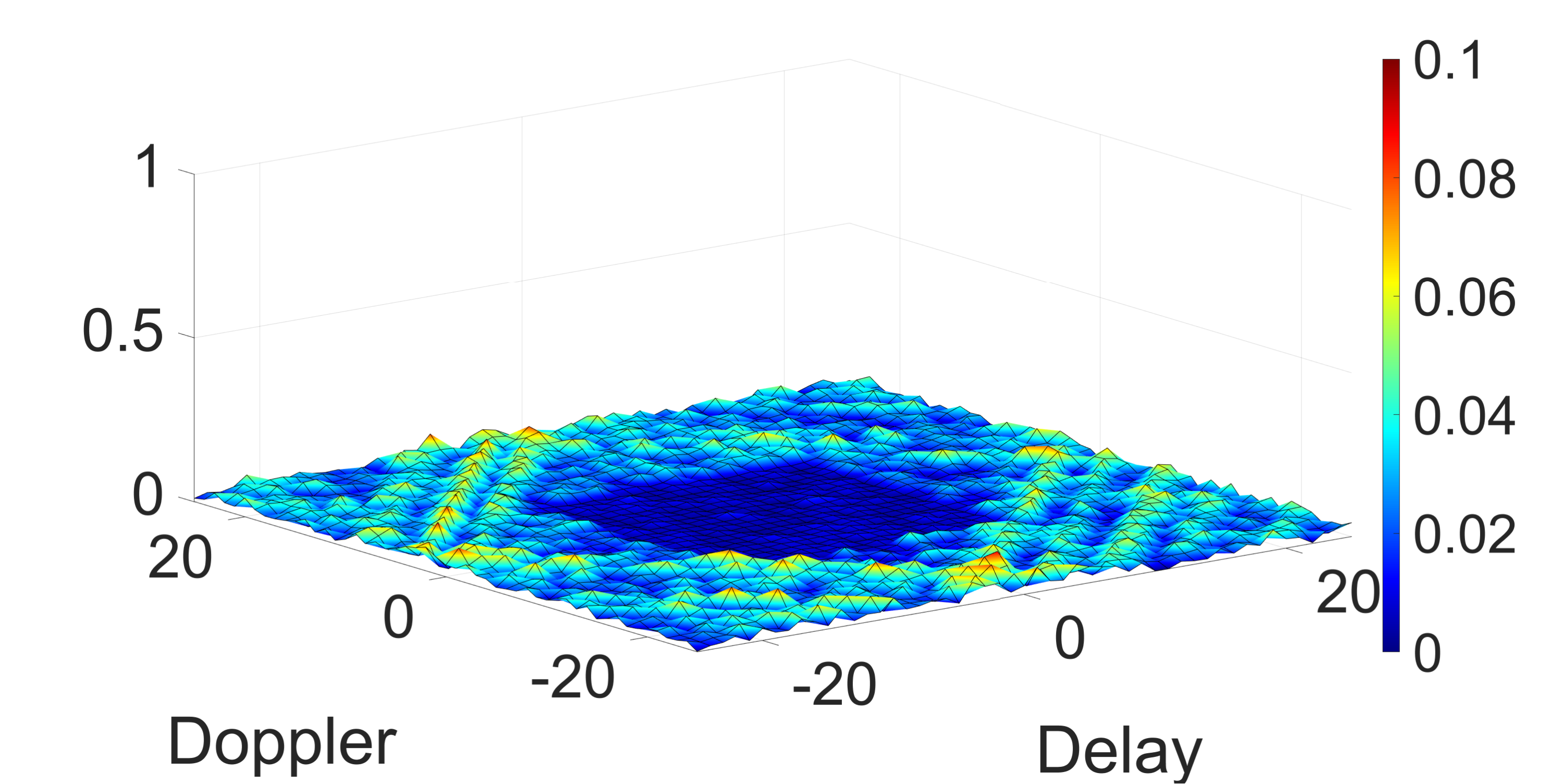}} \\ \footnotesize{(c)}\\
            
            {\includegraphics[trim=2.7cm 0cm 2.5cm 1.4cm, clip = true, width=1\linewidth]{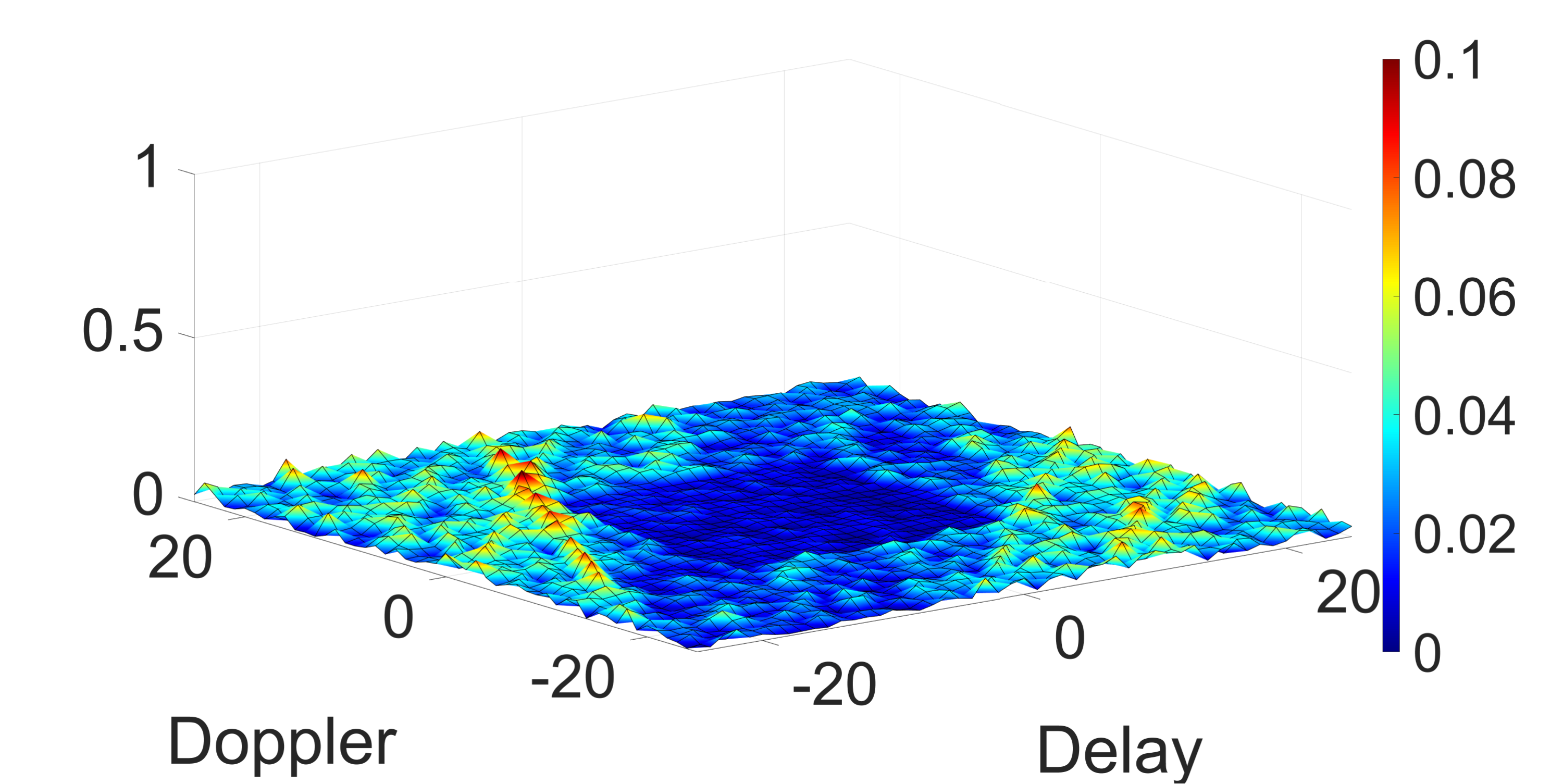}} \\ \footnotesize{(f)}\\
            
            {\includegraphics[trim=2.7cm 0cm 2.5cm 1.4cm, clip = true, width=1\linewidth]{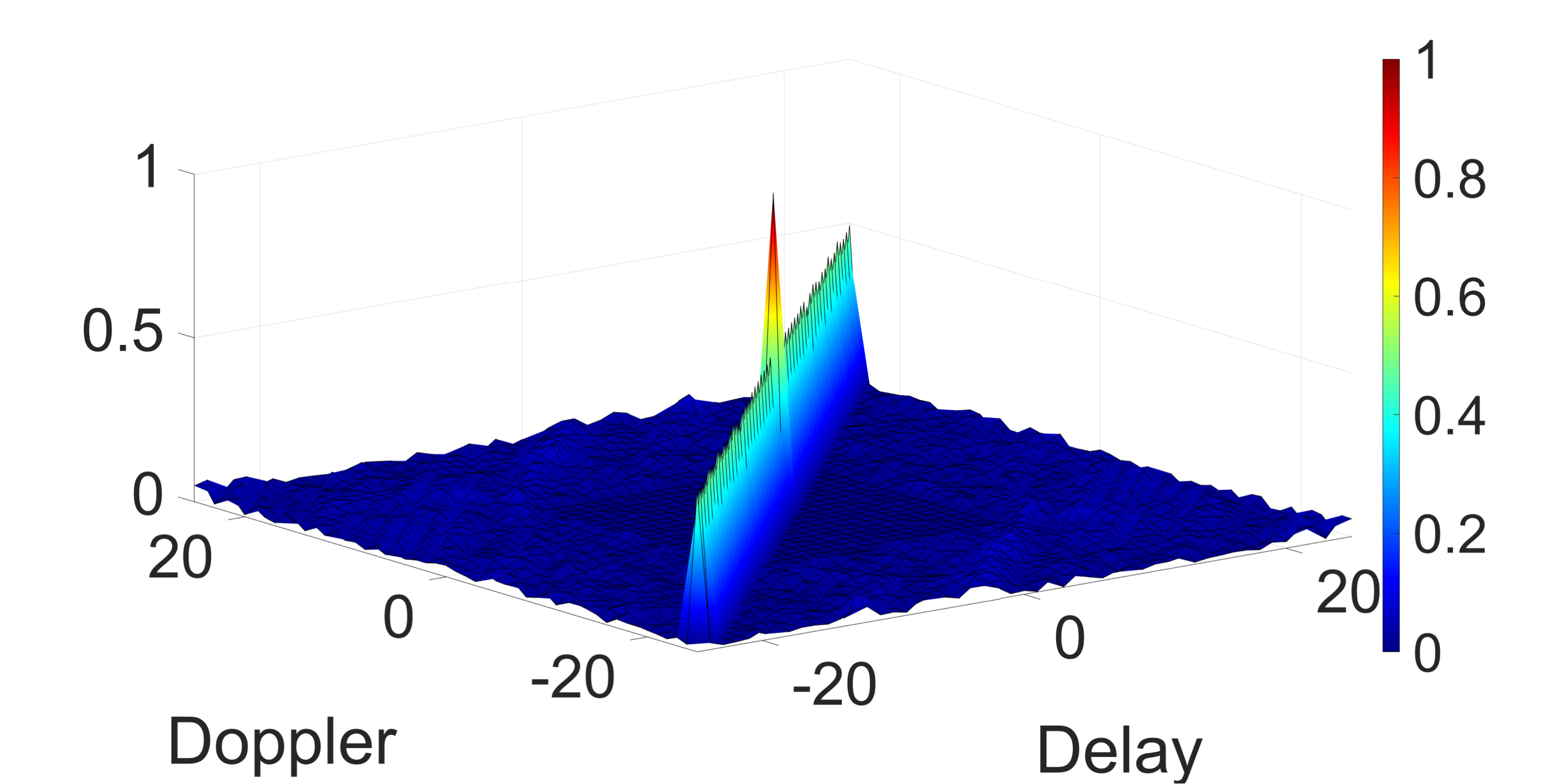}} \\ \footnotesize{(i)}\\
	\end{minipage}
        \begin{minipage}[b]{1.95\columnwidth}
		\centering
            \small{Illustration of an asymmetric Flag sequence set designed using Algorithm~1:}
            \footnotesize{$\bm f_{1}^{s} =[0.0215 + 0.0221j; 0.0185 + 0.0217j; 0.0203 + 0.0216j; 0.0438 - 0.0052j; 0.0396 + 0.0122j; \cdots]$}, \\ \footnotesize{$\bm f_{1}^{r} =[ 0.0140 + 0.0206j; 0.0126 + 0.0199j;0.0137 + 0.0201j;0.0444 - 0.0009j;0.0353 + 0.0166j;\cdots]$};\\
            \footnotesize{$\bm f_{2}^{s} =[0.0273 - 0.0217j;0.0268 - 0.0216j;0.0355 - 0.0171j;0.0228 - 0.0206j;0.0004 + 0.0075j; \cdots]$}, \\ \footnotesize{$\bm f_{2}^{r} =[0.0212 - 0.0222j; 0.0237 - 0.0220j;0.0317 - 0.0193j; 0.0192 - 0.0204j;-0.0001+0.000j;\cdots]$};\\
            \footnotesize{$\bm f_{3}^{s} =[0.0020 - 0.0091j;0.0124 + 0.0201j;0.0295 + 0.0215j;0.0088 + 0.0186j;0.0252 - 0.0207j; \cdots]$}, \\ \footnotesize{$\bm f_{3}^{r} =[0.0043 - 0.0131j;0.0209 + 0.0223j; 0.0354 + 0.0182j;0.0173 + 0.0224j;0.0351 - 0.0166j;\cdots]$}.
	\end{minipage}
	\caption{Periodic AFs of the proposed Flag sequence set (asymmetric transmit sequences and receive reference sequences); (a) $\bm A_{\bm f_{1}^{s},\bm f_{1}^{r}}$; (b) $\bm A_{\bm f_{1}^{s},\bm f_{2}^{r}}$; (c) $\bm A_{\bm f_{1}^{s},\bm f_{3}^{r}}$; (d) $\bm A_{\bm f_{2}^{s},\bm f_{1}^{r}}$; (e) $\bm A_{\bm f_{2}^{s},\bm f_{2}^{r}}$; (f) $\bm A_{\bm f_{2}^{s},\bm f_{3}^{r}}$; (g) $\bm A_{\bm f_{3}^{s},\bm f_{1}^{r}}$; (h) $\bm A_{\bm f_{3}^{s},\bm f_{2}^{r}}$; (i) $\bm A_{\bm f_{3}^{s},\bm f_{3}^{r}}$.}
    \label{AF_k3_PFS}
\end{figure*}

The AAFs and CAFs of the proposed Flag sequence set with $M=3$ and sequence length 1021 are presented in Fig. {\ref{AF_k3_PFS}}. Here, we use the AP-MM algorithm for the asymmetric transmit sequences and receive reference sequences with $\epsilon = 1$, and the ZoO is set to $\bm \Gamma \left (10,10 \right )$. We use three random sequences as the initial Peak sequences. The initial Curtain sequences are set with $\xi_1 = -1$, $\xi_2 = 2$, $\xi_3 = 1$ and $q_1 = 1$, $q_2 = 0$, $q_3 = 1$, satisfying Theorem~\ref{T1} and Corollary~\ref{C1}. In Fig. {\ref{AF_k3_PFS}}(a), (e), (i) $m_1=m_2$, one can see that the AAFs within the ZoO approaches an ideal \emph{peak-curtain} shape with low sidelobes. In the other sub-figures of Fig. {\ref{AF_k3_PFS}} with $m_1 \neq m_2$, corresponding to the CAFs of the sequence set, the sidelobe level within the ZoO is close to $0$. Such results demonstrate that each user can achieve low-complexity delay-Doppler estimation by utilizing the Flag method with our proposed Flag sequence sets. 

\begin{figure}[bp]
  \centering
  \begin{minipage}[b]{0.49\linewidth}
    \centering
    \includegraphics[trim=0.36cm 0cm 0.5cm 0.5cm, clip = true,width=\linewidth]{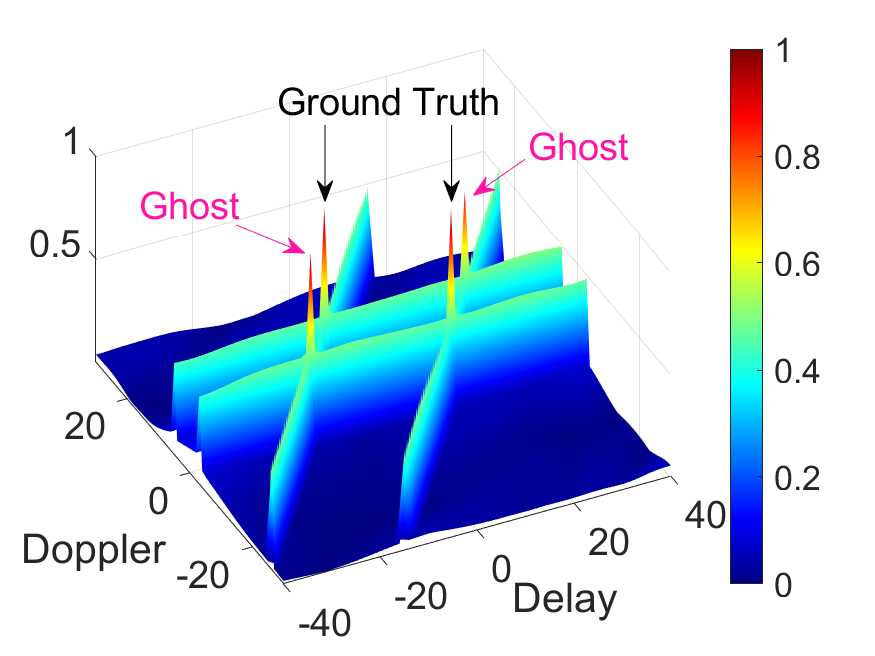}
    \footnotesize{(a)}
  \end{minipage}
  \hfill
  \begin{minipage}[b]{0.49\linewidth}
    \centering
    \includegraphics[trim=0.36cm 0cm 0.5cm 0.5cm, clip = true,width=\linewidth]{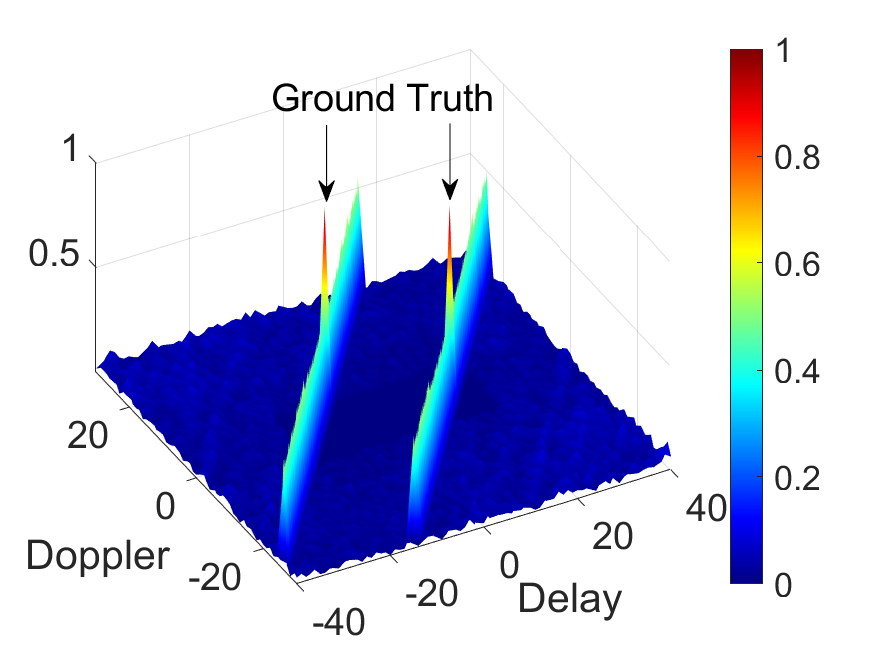}
    \footnotesize{(b)}
  \end{minipage}
  \caption{Periodic AAFs associated with 2 radar targets; (a) The HWS (with diagonal line, diagonal torus and $31$ as generator \cite[Section IV]{Fish2013}); (b) The proposed Flag sequence (symmetric transmit sequence and receive reference sequence: $\bm f^{s} =\bm f^{r} =[0.0279 + 0.0215i;	0.0421 - 0.0096i\cdots]$).}
  \label{AF_PFS_HWS} 
\end{figure}

Fig.~\ref{AF_PFS_HWS} compares the AAF of our proposed Flag sequence with that of the HWS \cite{Fish2013} of length 1021 in a multi-target scenario. In this example, we use the AP-MM algorithm for symmetric transmit sequence and receive reference sequence case with $M=1$ to generate our sequence, the ZoO is set to $\bm \Gamma \left (40,10 \right )$. The initial Peak sequence is the Weil sequence used in the HWS in Fig.~\ref{AF_PFS_HWS}(a), while the Curtain sequence uses $\xi = q = 1$. Two targets are set at ($-9$, $5$) and ($12$, $-3$) without loss of generality. It can be observed that our proposed sequence avoids high sidelobes and closely approximates two easily distinguishable \emph{peak-curtain} shapes in the AAF. In contrast, the high sidelobes generated by Weil sequences in the AAF of HWS can superimpose in multi-target scenarios, leading to ghost targets.

\begin{figure}[tbp]
  \centering
    {\includegraphics[width=0.99\linewidth]{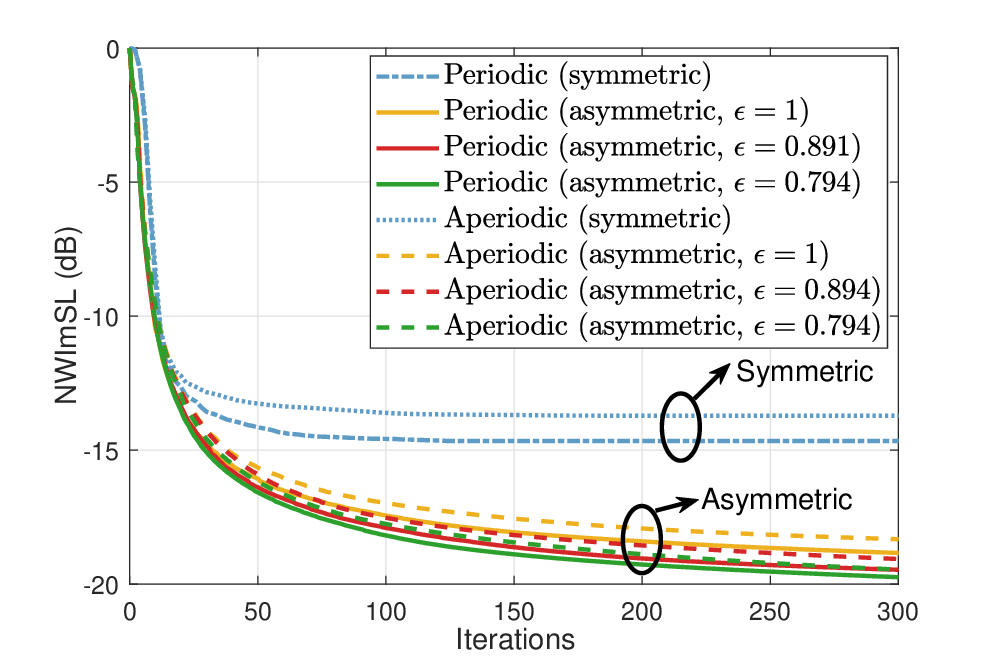}} 
  \caption{Evolution curves of the normalized Weighted Integrated masked Sidelobe Level (NWImSL) of our proposed algorithms.}
  \label{NSCWISL} 
\end{figure}

Fig.~\ref{NSCWISL} presents the evolution curves of the NWImSL with respect to the number of iterations for the AP-MM algorithms under both symmetric and asymmetric receive reference sequences cases. We set $M=3$, the sequence length is 509, and the ZoO is $\bm \Gamma \left (10,10 \right )$. To facilitate comparison, we use $3$ Weil sequences as the initial Peak sequences, while the Curtain sequences are set with $\xi_1 = q_1 = -1$, $\xi_2 = q_2 = 2$ and $\xi_3 = q_3 = 1$. The evolution curves demonstrate that the objective function monotonically decreases and gradually converges for both aperiodic and periodic cases, which proves that the AP-MM algorithms can effectively reduce the NWImSL of the sequence sets. Compared to the AP-MM algorithm for symmetric transmit sequences and receive reference sequences, the AP-MM algorithm for asymmetric case can achieve lower NWImSL due to its greater sequence design degree-of-freedom. One can also see that the performance of the AP-MM algorithm in the aperiodic case is slightly inferior to that in the periodic case, due to the extended receive reference sequences of the Curtain sequences part in the aperiodic case. 

Table \ref{CWISL and PSL Comparison} presents the NWImSL, PMmSR and PAPR for the HWS \cite{Fish2013} and the proposed Flag sequences. It also shows $\Delta$ in (\ref{assert}) and the LPG caused by the receiver reference sequences' extension and/or asymmetry. The sequence length and initial sequence here follow the settings of Fig.~\ref{NSCWISL}. WImSL of an HWS is used as the reference to calculate the NWImSL. The results reveal that our proposed Flag sequences exhibit significantly better NWImSL compared to HWS. Furthermore, the proposed Flag sequences demonstrate a notably higher PMmSR than HWS, indicating their closer proximity to the \emph{peak-curtain} AAF. It is worth noting that under the asymmetric case, selecting smaller values of $\epsilon$ does not lead to a higher PMmSR, despite yielding lower NWImSL. This behavior arises due to the LPG, which diminishes the amplitude of the mainlobe. Consequently, choosing a small value of $\epsilon$ is unnecessary. It can be observed that the PAPR of the proposed sequences is very close to the theoretical value of 3.01~dB (linear scale value of 2, see Section~\ref{Section3C}). The PAPR of HWS is also close to this value. Note that the HWS in Table~\ref{CWISL and PSL Comparison} does not include the high PAPR type constructed using the first type of Heisenberg sequence (Delta function, see Remark~\ref{remark1}). Moreover, we compare the average LPG with the theoretical value. Since the Peak sequences and Curtain sequences are not perfectly orthogonal, and the penalty function is an approximate constraint, there exists some deviation between the actual LPG and the theoretical value, although they are generally close to each other. As $M$ increases, it becomes more challenging for AP-MM algorithms to achieve perfect orthogonality, resulting in larger deviations at $M=3$ than at $M=1$. If stricter limits on LPG are desired, smaller $\beta$ should be used to increase the influence of the penalty. Additionally, the assertion in (\ref{assert}) is substantiated through numerical examples of $\Delta$ presented in this table. By substituting the proposed sequences generated by AP-MM algorithms into (\ref{assert}), we demonstrate the approximate orthogonality between $\bm{p}_m^s$ and $\bm{c}_m^s$, as well as between $\bm{p}_m^r$ and $\bm{c}_m^r$. This further validates the effectiveness of the constraints employed in Section~\ref{conssubs}.

\begin{table*} [t!]
	\centering
	\caption{Comparison of NWImSL and PMmSR in ZoO of the proposed Flag sequences and the HWS}
    \resizebox{2.03\columnwidth}{!}{
	\begin{tabular}{|c | c| c |  c| c | c|c|  c| c |c |c |c |}
		\hline
		\multirow{2}*{Sequences} & \multicolumn{2}{ c |}{NWImSL (dB)}& \multicolumn{2}{ c |}{PMmSR (dB)} & \multicolumn{2}{ c |}{PAPR (dB)} & \multicolumn{3}{ c |}{LPG (dB)} & \multicolumn{2}{ c |}{$\Delta$ as per (\ref{assert}) (dB)}\\
		\cline{2-12}
		~ & $M=1$ & $M=3$ & $M=1$ & $M=3$ & $M=1$ & $M=3$ & Theoretical &$M=1$ & $M=3$  &$M=1$ & $M=3$\\
		\hline
		HWS (periodic, symmetric) & $0$ & $0$ & $3.615$ & $3.615$ &  $2.990$ & $3.023$ &$-$ & $-$& $-$ &$-$& $-$\\
		\hline
		Proposed (periodic, symmetric) & $-47.356$ & $-14.659$ & $28.283$ & $11.758$& $2.984$ & $2.990$ & $-$& $-$& $-$ &$-19.581$& $-12.533$\\
		\hline
		Proposed (periodic, asymmetric, $\epsilon = 1$)  & $-54.916$ & $-19.305$ & $31.011$ & $13.368$& $2.980$ & $3.005$ & $2.988$& $-0.004$ & $-0.061$&$-18.762$& $-14.512$\\
		\hline
		Proposed (periodic, asymmetric, $\epsilon = 0.894$)  & $-57.991$ & $-19.901$ & $35.837$ & $12.304$& $2.991$ & $2.998$ & $-0.5$& $-0.487$ & $-0.526$&$-19.642$& $-13.018$\\
            \hline
		Proposed (periodic, asymmetric, $\epsilon = 0.794$)  & $-66.959$ & $-20.292$ & $38.553$ & $13.029$& $2.986$ & $3.004$ & $-1$& $-0.936$ & $-0.991$&$-18.945$& $-13.954$\\
		\hline
            Proposed (aperiodic, symmetric) & $-26.342$ & $-13.714$ & $11.275$ & $5.972$& $2.989$ & $2.989$ & $-0.085$& $-0.085$& $-0.096$&$-22.674$& $-19.509$ \\
		\hline
		Proposed (aperiodic, asymmetric, $\epsilon = 1$)  & $-30.956$ & $-18.698$ & $13.131$ & $6.184$& $2.995$ & $2.997$ & $-0.085$& $-0.011$ & $-0.155$&$-23.266$& $-18.346$\\
		\hline
		Proposed (aperiodic, asymmetric, $\epsilon = 0.894$)  & $-34.176$ & $-19.606$ & $13.348$ & $6.132$ & $2.989$ & $3.003$ & $-0.585$& $-0.573$ & $-0.614$&$-22.891$& $-17.682$\\
            \hline
		Proposed (aperiodic, asymmetric, $\epsilon = 0.794$)  & $-35.043$ & $-19.907$ & $12.320$ & $6.134$ & $2.999$ & $2.996$ & $-1.085$& $-1.036$ & $-1.091$&$-23.482$& $-18.247$\\
		\hline
	\end{tabular}
    }
	\label{CWISL and PSL Comparison}
\end{table*}
\subsection{Detection and Estimation Performances}
In this subsection, we analyze the detection and estimation performances of the proposed Flag sequences by using the Flag method and compare it with that of HWS \cite{Fish2013}. Our simulations were conducted in a millimeter-wave system operating at a carrier frequency of $f_{\text{cr}}=77$ GHz. Unless otherwise stated, we consider a practical scenario where the signal is limited to a bandwidth $B$, with the range and velocity of the targets being randomly distributed in $[-75, 75]\, \text{m}$ and $[-150, 150]\, \text{m/s}$, respectively. The continuous baseband transmit signal with bandwidth $B$ can be represented as \cite{Das2020}:
\begin{equation}
    s(t) = \sum_{n=0}^{N-1} f^{s}[n] \text{sinc}\left ( \pi B\left ( t-\frac{n}{B}  \right )  \right ), 
\end{equation}
where the $\text{sinc}$ function is defined as $\text{sinc}(x)=\sin(x)/x$. Then, the received signal can be represented as
\begin{equation}
    g(t) = \rho s(t-\tau) \text{exp} \left ( -j \varepsilon_c \omega(t-\tau)/N \right ) + z(t), 
\end{equation}
where $\rho$ is an unknown complex amplitude and $\varepsilon_c = 2 \pi f_{\text{cr}}$. $z(t)$ is the noise term, which is assumed to be an IID complex gain additive white Gaussian random process. Thus the distribution of noise after sampling is assumed to be $\mathcal{N} (0, \sigma_{z}^2)$. Then the SNR of the received signal can be calculated as $SNR = \rho^2/\sigma_{z}^2$. In the simulations, we set the bandwidth $B=10 \text{ MHz}$. For convenience in running $10^6$ Monte Carlo simulations, we used sequences of length 509 identical to those in Fig.~\ref{NSCWISL}. In practical applications, longer sequences can be chosen to further improve Doppler resolution and estimation performance.
\subsubsection{Detection Rate and False Alarm Rate}

\begin{figure}[htbp]
   \centering
    {\includegraphics[width=0.98\linewidth]{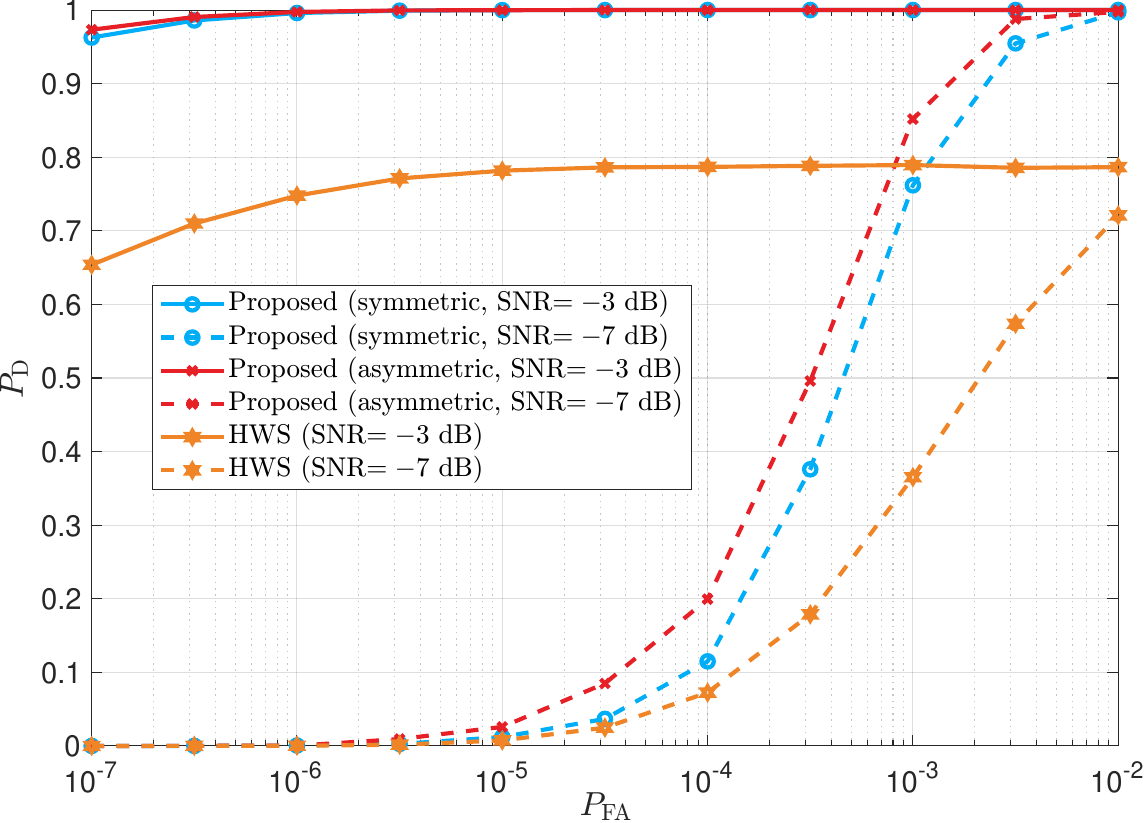}}
  \caption{Receiver operating characteristic (ROC) curves of the proposed Flag sequences and the HWS \cite{Fish2013}.}
  \label{ROC} 
\end{figure}


Fig.~\ref{ROC} shows the receiver operating characteristic (ROC) curves of our proposed Flag sequences and HWS \cite{Fish2013} with the Flag method. A false alarm is declared if an un-transmitted Flag sequence is detected. We used a constant false alarm rate (CFAR) detector with detection threshold \mbox{$\eth  =  -2\sigma_{z}^2\log\left(P_{\rm FA}\right)$}, where $P_{\rm FA}$ is the false alarm rate \cite{Richards2005}. If a point exceeding the threshold $\eth$ is identified during the first linear search of the Flag method, a second linear search is conducted along the \emph{curtain} direction. At this step, if a point is found that exceeds the sum of the threshold $\eth$ and the \emph{curtain} amplitude (approximated by the average amplitude of the line in the second linear search), it is classified as a target. The detection rate $P_{\rm D}$ represents the probability of successfully detecting targets within the error margins of one delay/Doppler bin using the Flag method, without detecting non-existent targets. For brevity, we apply the AP-MM algorithm for asymmetric transmit sequence and receive reference sequence with $\epsilon=1$ in Fig.~\ref{ROC}. Obviously, HWS exhibits inferior performance compared to our proposed Flag sequences. This performance difference is primarily due to the high AF sidelobes except for the \emph{peak-curtain} of the HWS (see Fig.~\ref{AF_PFS_HWS}(a)), which can lead to ghost targets. Additionally, the results indicate that the detection performance of the proposed Flag sequences designed for the asymmetric case is marginally superior to that of the sequences designed for the symmetric case. This matches with the results shown in Fig.~\ref{NSCWISL} and Table~\ref{CWISL and PSL Comparison}, indicating that the asymmetric transmit sequences and receive reference sequences with lower sidelobe levels lead to improved detection performance. Therefore, when the system and hardware conditions allow, using asymmetric transmit sequences and receive reference sequences can provide detection performance benefits.

\begin{figure}[tbp]
   \centering {\includegraphics[trim=0.9cm 0.1cm 0.8cm 0.75cm, clip = true, width=0.98\linewidth]{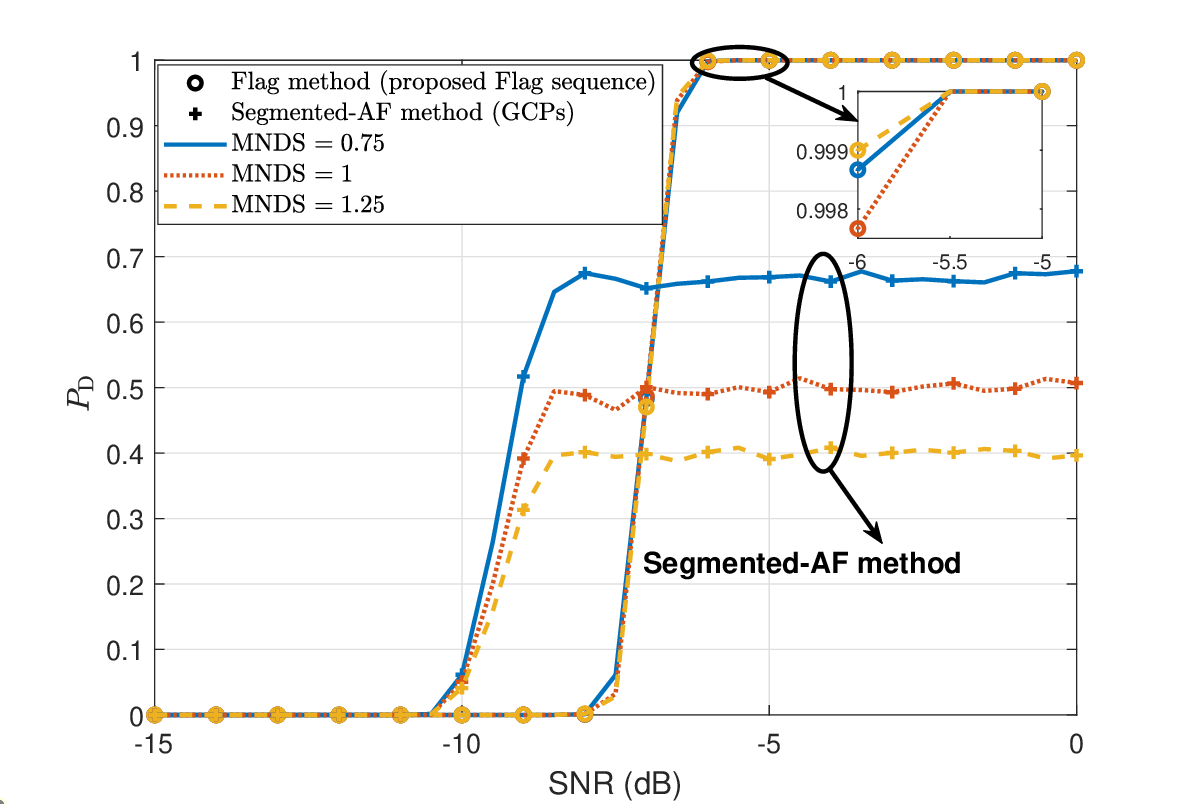}}
  \caption{Detection rate $P_{\text{D}}$ of Flag Method using proposed Flag sequence and Segmented-AF Method \cite{Zeng2020,Kumari2018} using GCPs.}
  \label{DRS} 
\end{figure}

Fig.~\ref{DRS} compares the detection rate of the Flag method using the proposed Flag sequence with the segmented-AF method \cite{Zeng2020,Kumari2018} using the Golay complementary pairs (GCPs), under $P_{\rm FA}=10^{-5}$. The segmented-AF method refers to the concept of multi-pulse radar, where a long sequence is divided into multiple segments. By assuming that the Doppler effect within a pulse/segment is negligible, the 2D-FFT is used to achieve an $\mathcal{O}(KN\log(N))$ real-time processing complexity for the exhaustive search of the 2D segmented-AF, which is higher than that of the Flag method ($\mathcal{O}(N\log(N))$, see Section~\ref{FMCC}). The segmented-AF method and its assumptions require the maximum normalized Doppler of one segment (MNDS in short, equivalent to $\omega_{\text{max}}$ divided by the number of segments, where $\omega_{\text{max}}$ is defined in Section~\ref{ZoODF}) to be less than or equal to $0.5$. This condition often imposes high signal bandwidth requirements. Based on the IEEE 802.11ad preamble considered in \cite{Zeng2020, Kumari2018}, we tested the segmented-AF method using GCPs consisting of $8$ segments, each with a length of $256$. For the Flag method, we use the proposed Flag sequence of the same length of $8 \times 256=2048$. As shown in Fig.~\ref{DRS}, when the MNDS exceeds $0.5$, the segmented-AF method fails to achieve good detection rate, whereas the Flag method remains stable for different MNDS. This suggests that the segmented-AF method using non-Flag sequences does not work well when the signal bandwidth is limited (resulting in MNDS greater than 0.5), while the proposed Flag Method using Flag sequence is not subject to this limitation.

\subsubsection{Mean Squared Error of Delay-Doppler Estimation}
Furthermore, we analyze the estimation performance of the proposed Flag sequences. The lower bound of MSE is an important criterion for evaluating the performance of delay-Doppler estimation. In this regard, the CRLB and SB are commonly considered \cite{Richards2005}. The CRLB of $\bm \eta = [\tau, \omega]^{\text{T}}$ for a Gaussian conditional observation is given by the inverse of the Fisher information matrix (FIM) \cite{Dogandzic2001}
\begin{equation}
    \text{CRLB}_{\bm \eta} = \frac{1}{2\text{SNR}} {\Re \left \{\bm \Phi_{\bm \eta}\right \}}^{-1}.
\end{equation}
Following the results presented in \cite{Das2020}, we have 
\begin{equation}
         {\Re \left \{\bm \Phi_{\bm \eta}\right \}} =\begin{bmatrix}
 \phi_{1,1} & \phi_{1,2} \\
  \phi_{2,1}&\phi_{2,2} 
\end{bmatrix},
\end{equation}
where 
\begin{subequations}
    \begin{align}
        &\phi_{1,1} = B^2\left ( \bm s^{\dagger } \bm H \bm s - \frac{{\left | \bm s^{\dagger } \bm T \bm s  \right |}^2 }{\bm s^{\dagger } \bm s }  \right ),\\
        &\phi_{1,2} = \phi_{2,1} = \varepsilon_{c} \Im \left \{  \bm s^{\dagger } \bm D \bm T \bm s - \frac{ \bm s^{\dagger } \bm D \bm s \bm s^{\dagger } \bm T \bm s  }{\bm s^{\dagger } \bm s }\right \}, \\
        &\phi_{2,2} = \frac{\varepsilon_{c}^2}{B^2}\left (  \bm s^{\dagger } \bm D^2 \bm s - \frac{{\left ( \bm s^{\dagger } \bm D \bm s  \right )}^2 }{\bm s^{\dagger } \bm s }\right ),
    \end{align}
\end{subequations}
with 
\begin{equation*}
    \begin{split}
        &\bm D = \text{Diag} \left( [N_1, N_1+1, \cdots, N_2] \right),\\
        &H[m,n]=\begin{cases}
	\frac{\pi^2}{3} , & m = n; \\
	\left ( -1 \right ) ^{\left | m-n \right | } \frac{2}{{\left ( m-n \right ) }^2}, & m \neq n;\\
        \end{cases}\\
        &T[m,n]=\begin{cases}
	0 , & m = n; \\
	 \frac{\left ( -1 \right ) ^{\left | m-n \right | }}{{\left ( m-n \right ) }}, & m \neq n,\\
        \end{cases}
    \end{split}
\end{equation*}
for samples with $N_1\le 0$, $N_2 \ge N$.

The SB is the lower limit of resolution due to the sampling rate limitation \cite{Richards2005}. The SB in term of range and velocity can be expressed as
$\text{SB}_{\text{range}} = \frac{c^2}{48 B^2  k_{\tau}^2}$ and $\text{SB}_{\text{speed}} = \frac{c^2 B^2}{48 f_{\text{cr}}^2 N^2 k_{\omega}^2}$, where $k_{\tau}$ and $k_{\omega}$ are the oversampling rates in time domain and frequency domain respectively.
\begin{figure}[tb!]
  \centering
    {\includegraphics[ width=0.95\linewidth]{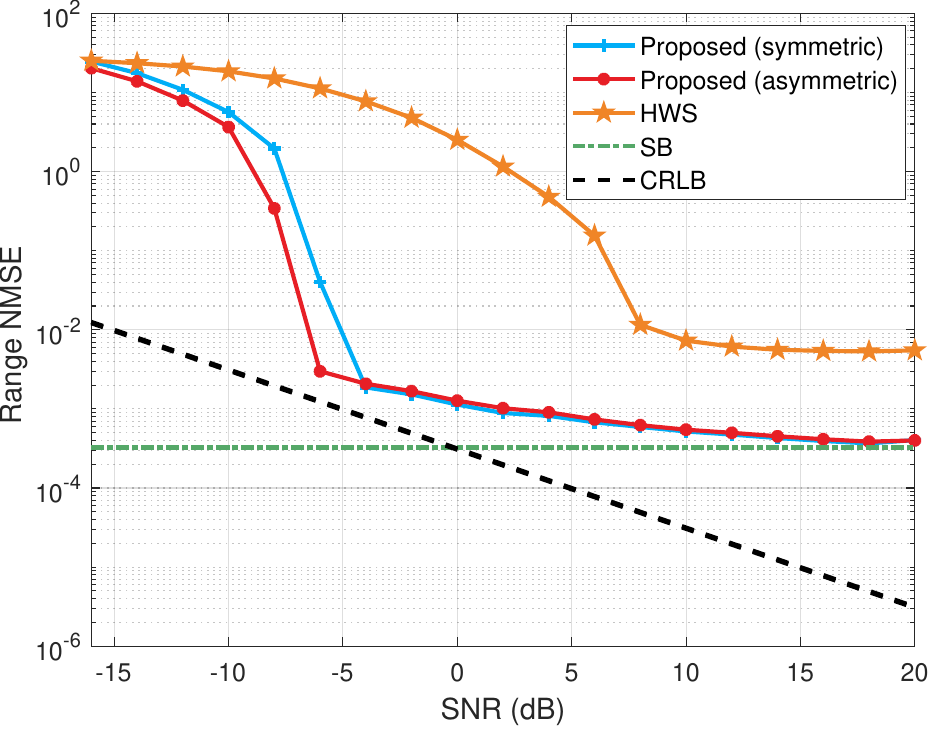}} \\ \footnotesize(a)\\
    {\includegraphics[ width=0.95\linewidth]{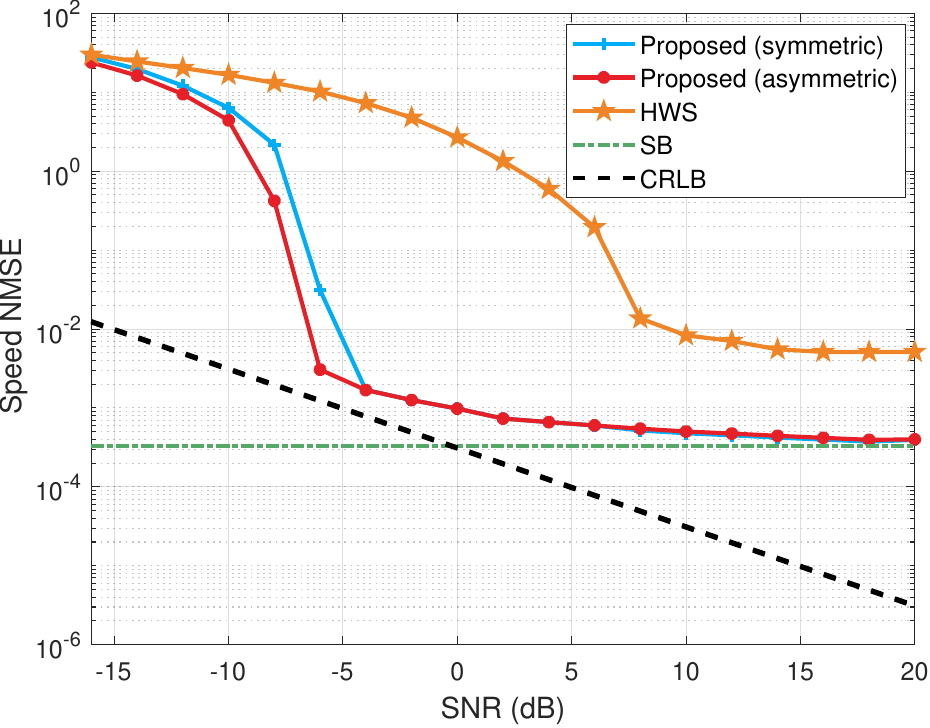}} \\ \footnotesize(b)\\
  \caption{Normalized estimation Mean Squared Errors (MSE), Cramér-Rao Lower Bound (CRLB) and Sampling Bound (SB); (a) Range; (b) Speed.}
  \label{MSE_os} 
\end{figure}

Fig.~\ref{MSE_os} illustrates a comparison of the range and speed MSE for our proposed Flag sequences and HWS \cite{Fish2013}. CRLB and SB are also provided as benchmarks. The normalized MSE (NMSE) are normalized based on the delay/Doppler bin sizes, respectively. In the simulations, we first estimate the integer delay and Doppler and then estimate the fractional ones using oversampling rates of $k_{\tau} = k_{\omega} = 16$. The results demonstrate that the proposed Flag sequences for the asymmetric case exhibit lower range and speed MSE compared to those for the symmetric case, implying improved detection and estimation performances due to their lower NWImSL (See Fig.~\ref{NSCWISL} and Table~\ref{CWISL and PSL Comparison}). Moreover, the high sidelobe level of HWS results in inferior estimation performance under the Flag method, as neither its range MSE nor its speed MSE can approach the CRLB or the SB. In contrast, our proposed Flag sequences demonstrate MSE that closely approximates the CRLB and SB at high SNR. Hence, our proposed Flag sequences possess significantly improved estimation performance compared to the HWS.

\subsection{Challenges and Open Problems on the Applications of Flag Sequences}
The proposed Flag sequences provide higher flexibility in sequence length and achieve lower sidelobes compared to the HWS \cite{Fish2013}, thus leading to improved detection and estimation performance. Despite these advantages, practical challenges persist. This subsection outlines these challenges and discusses open problems for future research.

The design and analysis of the Flag sequences in this paper and \cite{Fish2013} are based on the assumption that each target can be approximated as a single point target, which is a common simplification in many studies. However, in practical scenarios, some targets may be extended objects or generate a large number of scattering points. In such cases, the AAF of Flag sequences may exhibit extended or distorted \emph{curtains}, potentially complicating detection and estimation. Moreover, clutter arising from undesired reflections in the environment (e.g., ground, rain, buildings, etc.) may also obscure weak targets. Conventional radar sequence design often mitigates clutter through adaptive filtering, clutter cancellation, or spatial processing \cite{9109917,9564776}. Given the presence of the \emph{curtain} in the AAF, clutter mitigation for Flag sequences may be more challenging. Future research may investigate adaptive sequence optimization methods and explore advanced clutter removal techniques for Flag sequences to enhance their applicability in extended-target and clutter-rich scenarios.  

Additionally, the \emph{curtain} of Flag sequences may influence weak target detection under certain conditions. Specifically, when transmitting a Flag sequence designed based on a \emph{curtain} sequence $\bm c_{\xi,q}$, we consider a strong target with delay $\tau_{\text{s}}$ and Doppler shift $\omega_{\text{s}}$, and a weak target with delay $\tau_{\text{w}}$ and Doppler shift $\omega_{\text{w}}$. If their velocities $v_{\text{s}}, v_{\text{w}}$ and ranges $d_{\text{s}}, d_{\text{w}}$ satisfy $\xi B^2(d_{\text{w}} - d_{\text{s}}) = N f_{\text{cr}} (v_{\text{w}} - v_{\text{s}})$, meaning that $\omega_{\text{w}} - \omega_{\text{s}} = \xi(\tau_{\text{w}} - \tau_{\text{s}})$, the \emph{curtains} of the two targets will overlap. In this case, the strong target's \emph{curtain} may obscure the weak target's \emph{peak} depending on their relative phases, leading to potential missed detections in single-frame processing. Since the relative ranges and velocities of different targets evolve over time, possible solutions to this issue include mitigating it through tracking or multiple detections, or leveraging previous detection results to predict potential overlaps.

Overall, the proposed Flag sequences offer significant advantages over HWS under the point target assumption but may still face challenges in scenarios with extended or targets, clutter, and occasional obscuration of weak targets by the \emph{curtains} of strong targets. To ensure its effectiveness, it is preferred to have scenarios where the point target assumption approximately holds, with a priority on detecting strong targets or the ability to detect weak targets through tracking techniques.

\section{Conclusions}
In this paper, we have designed improved Flag sequence sets for low-complexity delay-Doppler estimation in modern communication and radar systems. Traditional Flag sequences, i.e., HWS, are limited to prime lengths and consider only periodic AF with symmetric receive reference sequences. By contrast, our proposed method leads to Flag sequences of arbitrary lengths for both periodic and aperiodic AFs with symmetric/asymmetric receive reference sequences. We have first proposed novel theorems and corollaries for systematic constructions of Curtain sequence sets with ideal \emph{curtain} AAFs and zero/near-zero CAFs in ZoO. The connections between Curtain sequences and the parameter selection for chirp sequences are elaborated. We have then developed AP-MM algorithms which can efficiently minimize WImSL by jointly optimizing the transmit Flag sequences and symmetric/asymmetric receive reference sequences. Simulation results demonstrate that our proposed Flag sequences outperform HWS in terms of WImSL, PMmSR, and \emph{peak-curtain} AAF, leading to improved detection and estimation performances. Additionally, it is shown that our proposed Flag sequences with asymmetric receive reference sequences perform the best due to larger sequence design degree-of-freedom. Moreover, the MSE of our proposed Flag sequences under the low-complexity Flag method approach the CRLB and SB.

Future research will focus on improving the robustness of Flag sequences against specific effects such as clutter and speckle noise, as well as extending their applicability to extended targets. The proposed optimization framework will be extended to ISAC areas, and new algebraic constructions of Flag sequences will be explored.
\appendices
\section{Proof of Theorem~\ref{T1}}
\label{T1proof}
For any integer $u$, $0<u<N$, one can verify that \cite{Sarwate1979}
\begin{equation}
\label{sum0}
    \sum_{n=0}^{N-1} \text{exp} \left (\pm \frac{j2 \pi u r n}{N} \right )=0, \quad \text{exp} \left (\frac{j2 \pi r}{N} \right ) \neq 1,
\end{equation} 
where $r$ is any integer relatively prime to $N$. Then, for $\bm c_{\xi,q}$ that satisfies $[\xi N-q]_2 = 0$, its periodic AAF in a ZoO $\bm \Gamma$ with $\left|\xi \right|\tau_{\text{max}}+\omega_{\text{max}}<N$ can be calculated as:
\begin{align}
    &A_{\bm c_{\xi,q},\bm c_{\xi,q}}(\tau,\omega) \nonumber \\ 
     & = \frac{1}{N} \bigg |\sum_{n=0}^{N-\tau-1} \text{exp} \left (j\pi\frac{2\omega n-2\xi \tau n -\xi \tau^2-q\tau}{N} \right ) \nonumber \\
     & + \sum_{n=0}^{N-\tau-1} \text{exp} \left (j\pi\frac{2\omega n-2\xi \tau n -\xi \tau^2-q\tau}{N}-\left (\xi N-q \right ) \right ) \bigg | \nonumber \\
     & = \frac{1}{N} \left |\text{exp} \left (-j\pi\frac{\xi \tau^2+q\tau}{N} \right ) \sum_{n=0}^{N-1} \text{exp} \left (j\pi\frac{2(\omega -\xi \tau) n }{N} \right ) \right | \nonumber \\
     & = \begin{cases}
     	1,\quad & (\tau,\omega) \in \bm \Gamma  \text{ and } \omega = \xi \tau;\\
            0,\quad & (\tau,\omega) \in \bm \Gamma  \text{ and } \omega \neq \xi \tau.
         \end{cases}
         \label{CuAF}
\end{align}
Therefore, the periodic AAF of $\bm c_{\xi,q}$ that satisfies $[\xi N-q]_2 = 0$ exhibits an ideal \emph{curtain} along the line $\omega=\xi\tau$ with $0$ sidelobes elsewhere within the ZoO $\bm \Gamma$.

\section{Proof of Corollary~\ref{C1}}
\label{C1proof}
The squared periodic CAF of two Curtain sequences $\bm a$ and $\bm b$, obtained from Therorem~\ref{T1}, can be expressed as
\begin{align}
    \label{SQAF}
        &\left |A_{\bm a,\bm b}(\tau,\omega) \right|^2 \nonumber \\
         & =\! \sum_{t=0}^{N-1}\sum_{s=0}^{N-1}  a[t] \left (b[t\!+\!\tau]\right)^{\ast}  \left( a[s] \right )^{\ast} b[s\!+\!\tau]\text{exp} \left (\frac{j2 \pi \omega (t\!-\!s)}{N} \right)   \nonumber \\
         & = \frac{1}{N^2}\sum_{t=0}^{N-1} \sum_{s=0}^{N-1}  \text{exp} \bigg (\frac{j \pi}{N}((\xi_a-\xi_b)(t^2-s^2) \nonumber \\
         & \quad \quad \quad +(q_a-q_b)(t-s)+2(\omega -\xi_b \tau)(t-s)) \bigg ).
\end{align}
Let $e = t-s, e = 0,1,\cdots,N-1$. Then (\ref{SQAF}) is equal to
\begin{align}
\label{SQAF2} 
    &|A_{\bm a,\bm b}(\tau,\omega) |^2  \nonumber \\
         & = \frac{1}{N^2}\sum_{t=0}^{N-1} \sum_{e=0}^{N-1}  \text{exp} \bigg (\frac{j e \pi}{N}((\xi_a-\xi_b)(2t-e) \nonumber \\
         & \quad \quad \quad \quad \quad \quad \quad \quad \quad +(q_a-q_b)+2(\omega -\xi_b \tau)) \bigg)  \nonumber\\
         & = \frac{1}{N}+\frac{1}{N^2} \sum_{e=1}^{N-1}  \text{exp} \bigg (\frac{j e \pi}{N}(-(\xi_a-\xi_b)e +(q_a-q_b) \nonumber \\
         & \quad +2(\omega -\xi_b \tau)) \bigg )\sum_{t=0}^{N-1}\text{exp} \left(\frac{j2\pi e t (\xi_a-\xi_b)}{N} \right ).
\end{align}
When $\left |\xi_a-\xi_b \right |$ is relatively prime to $N$, according to (\ref{sum0}), the summation $\sum_{t=0}^{N-1}\text{exp} \left (\frac{j2\pi e t (\xi_a-\xi_b)}{N} \right )$ in (\ref{SQAF2}) is $0$ when $e \neq 0$. Thus, the CAF $|A_{\bm a,\bm b}(\tau,\omega) |=1/\sqrt{N}$ for any $(\tau,\omega)$. 

\section{Proof of Propositions~\ref{P1} and \ref{P2}}
\label{Ppproof}
The proofs of Propositions~\ref{P1} and \ref{P2} rely on the following lemma \cite{Song2016}. 
\begin{lemma} 
\label{lemma1}
    Let $\bm Y$ be a $d \times d$ Hermitian matrix and $\bm Z$ be another $d \times d$ Hermitian matrix such that $\bm Z \succeq \bm Y$. Then for any point $\bm x^{(t)} \in \mathbb{C}^{d}$, the following inequality holds
    \begin{align}
    \label{lemma1ineq}
        \bm x^{\dagger} \bm Y \bm x\!\le\! \bm x^{\dagger} \bm Z \bm x \!+\! 2 \Re \left \{\bm x^{\dagger}(\bm Y \!- \!\bm Z)\bm x^{(t)}\! \right \} \! +\! {\bm x^{(t)}}^{\dagger}(\bm Z\!-\! \bm Y) \bm x^{(t)}.
    \end{align}
    Thus, the quadratic function $\bm x^{\dagger} \bm Y \bm x$ of $\bm x$ can be majorized by the right-hand side of (\ref{lemma1ineq}) at the $t$th iteration $\bm x^{(t)}$.
\end{lemma}

\begin{IEEEproof}
   Lemma~\ref{lemma1} is proved in \cite{Song2016} and hence omitted here. For more details on the MM method, refer to \cite{Song2016}.
\end{IEEEproof}

Leveraging Lemma~\ref{lemma1}, one can identify suitable surrogate functions for the optimization problems (\ref{PA3}) and (\ref{PB2}), respectively. After ignoring the constant terms, we obtain (\ref{PA4}) of Proposition~\ref{P1} and (\ref{PB3}) of Proposition~\ref{P2}. In Proposition~\ref{P1}, $\lambda_{\text{max}}(\bm \Lambda)$ can be obtained by \cite{Song2016}
\begin{align*}
    &\lambda_{\text{max}}(\bm \Lambda) =\max \left \{{\lambda_{\text{max}}(\alpha \bm \Lambda_1),\lambda_{\text{max}}((1-\alpha)\bm \Lambda_2), \lambda_{\text{max}}(\beta' \bm \Lambda_3)}\right \},\\
    &\lambda_{\text{max}}(\alpha \bm \Lambda_1) = \alpha \max_{\tau} \left \{2L-2\tau | (\tau,\omega) \in \bm \Gamma  \right \}, \\
    &\lambda_{\text{max}}((1-\alpha) \bm \Lambda_2) = (1-\alpha) \max_{\tau} \left \{2L-2\tau | (\tau,\omega) \in \bm \Gamma  \right \}, \\
    &\lambda_{\text{max}}(\beta' \bm \Lambda_3) = \beta' L.
\end{align*}
In Proposition~\ref{P2}, we can choose $\widetilde{\lambda}$ as \cite{Wang2021}
\begin{align*}
    \widetilde{\lambda} &= 4ML \max_{1\le a,b \le 2ML} \left |\Omega_{\bm x^{(t)}}[a,b] + {\Omega_{\bm x^{(t)}}}^{\dagger}[a,b]\right| \nonumber \\
    &\ge \lambda_{\text{max}} \left( \bm \Omega_{\bm x^{(t)}} + {\bm \Omega_{\bm x^{(t)}}}^{\dagger} \right) \nonumber\\
    & >\lambda_{\text{max}} \left( \bm \Omega_{\bm x^{(t)}} + {\bm \Omega_{\bm x^{(t)}}}^{\dagger} - 2 \lambda \bm x^{(t)} {\bm x^{(t)}}^{\dagger} \right).
\end{align*}

\bibliographystyle{IEEEtran}

\end{document}